\documentclass[12pt]{article} 
\usepackage{amsmath,amssymb,amsbsy,dsfont,amsfonts,epsfig
           ,vmargin,fdag,color}

\definecolor{GreenYellow}{cmyk}{0.15,0,0.69,0}
\definecolor{Yellow}{cmyk}{0,0,1,0}
\definecolor{Goldenrod}{cmyk}{0,0.10,0.84,0}
\definecolor{Dandelion}{cmyk}{0,0.29,0.84,0}
\definecolor{Apricot}{cmyk}{0,0.32,0.52,0}
\definecolor{Peach}{cmyk}{0,0.50,0.70,0}
\definecolor{Melon}{cmyk}{0,0.46,0.50,0}
\definecolor{YellowOrange}{cmyk}{0,0.42,1,0}
\definecolor{Orange}{cmyk}{0,0.61,0.87,0}
\definecolor{BurntOrange}{cmyk}{0,0.51,1,0}
\definecolor{Bittersweet}{cmyk}{0,0.75,1,0.24}
\definecolor{RedOrange}{cmyk}{0,0.77,0.87,0}
\definecolor{Mahogany}{cmyk}{0,0.85,0.87,0.35}
\definecolor{Maroon}{cmyk}{0,0.87,0.68,0.32}
\definecolor{BrickRed}{cmyk}{0,0.89,0.94,0.28}
\definecolor{Red}{cmyk}{0,1,1,0}
\definecolor{OrangeRed}{cmyk}{0,1,0.50,0}
\definecolor{RubineRed}{cmyk}{0,1,0.13,0}
\definecolor{WildStrawberry}{cmyk}{0,0.96,0.39,0}
\definecolor{Salmon}{cmyk}{0,0.53,0.38,0}
\definecolor{CarnationPink}{cmyk}{0,0.63,0,0}
\definecolor{Magenta}{cmyk}{0,1,0,0}
\definecolor{VioletRed}{cmyk}{0,0.81,0,0}
\definecolor{Rhodamine}{cmyk}{0,0.82,0,0}
\definecolor{Mulberry}{cmyk}{0.34,0.90,0,0.02}
\definecolor{RedViolet}{cmyk}{0.07,0.90,0,0.34}
\definecolor{Fuchsia}{cmyk}{0.47,0.91,0,0.08}
\definecolor{Lavender}{cmyk}{0,0.48,0,0}
\definecolor{Thistle}{cmyk}{0.12,0.59,0,0}
\definecolor{Orchid}{cmyk}{0.32,0.64,0,0}
\definecolor{DarkOrchid}{cmyk}{0.40,0.80,0.20,0}
\definecolor{Purple}{cmyk}{0.45,0.86,0,0}
\definecolor{Plum}{cmyk}{0.50,1,0,0}
\definecolor{Violet}{cmyk}{0.79,0.88,0,0}
\definecolor{RoyalPurple}{cmyk}{0.75,0.90,0,0}
\definecolor{BlueViolet}{cmyk}{0.86,0.91,0,0.04}
\definecolor{Periwinkle}{cmyk}{0.57,0.55,0,0}
\definecolor{CadetBlue}{cmyk}{0.62,0.57,0.23,0}
\definecolor{CornflowerBlue}{cmyk}{0.65,0.13,0,0}
\definecolor{MidnightBlue}{cmyk}{0.98,0.13,0,0.43}
\definecolor{NavyBlue}{cmyk}{0.94,0.54,0,0}
\definecolor{RoyalBlue}{cmyk}{1,0.50,0,0}
\definecolor{Blue}{cmyk}{1,1,0,0}
\definecolor{Cerulean}{cmyk}{0.94,0.11,0,0}
\definecolor{Cyan}{cmyk}{1,0,0,0}
\definecolor{ProcessBlue}{cmyk}{0.96,0,0,0}
\definecolor{SkyBlue}{cmyk}{0.62,0,0.12,0}
\definecolor{Turquoise}{cmyk}{0.85,0,0.20,0}
\definecolor{TealBlue}{cmyk}{0.86,0,0.34,0.02}
\definecolor{Aquamarine}{cmyk}{0.82,0,0.30,0}
\definecolor{BlueGreen}{cmyk}{0.85,0,0.33,0}
\definecolor{Emerald}{cmyk}{1,0,0.50,0}
\definecolor{JungleGreen}{cmyk}{0.99,0,0.52,0}
\definecolor{SeaGreen}{cmyk}{0.69,0,0.50,0}
\definecolor{Green}{cmyk}{1,0,1,0}
\definecolor{ForestGreen}{cmyk}{0.91,0,0.88,0.12}
\definecolor{PineGreen}{cmyk}{0.92,0,0.59,0.25}
\definecolor{LimeGreen}{cmyk}{0.50,0,1,0}
\definecolor{YellowGreen}{cmyk}{0.44,0,0.74,0}
\definecolor{SpringGreen}{cmyk}{0.26,0,0.76,0}
\definecolor{OliveGreen}{cmyk}{0.64,0,0.95,0.40}
\definecolor{RawSienna}{cmyk}{0,0.72,1,0.45}
\definecolor{Sepia}{cmyk}{0,0.83,1,0.70}
\definecolor{Brown}{cmyk}{0,0.81,1,0.60}
\definecolor{Tan}{cmyk}{0.14,0.42,0.56,0}
\definecolor{Gray}{cmyk}{0,0,0,0.50}
\definecolor{Black}{cmyk}{0,0,0,1}
\definecolor{White}{cmyk}{0,0,0,0}

\setmarginsrb{2.5cm}{2.5cm}{2.5cm}{2cm}{0cm}{0cm}{1cm}{1cm}

\newcommand{\tr}{\mathrm{tr}}
\newcommand{\id}{\mathrm{d}}
\newcommand{\ii}{\mathord{\mathrm{i}}}

\newlength{\largfig}
\def\ds#1{#1\kern-1ex\hbox{/}} 
\def\sl#1{#1\kern-1ex\hbox{/}} 
\def\dsh{h\kern-1.2ex /}

\def\beq{\begin{equation}} 
\def\eeq{\end{equation}} 
\def\eq{\beq\eeq} 
\def\beqn{\begin{eqnarray}} 
\def\eeqn{\end{eqnarray}}

\def\({\left(} 
\def\){\right)} 
 
\def\ba{\begin{eqnarray}} 
\def\ea{\end{eqnarray}} 
\def\bq{\begin{equation}} 
\def\eq{\end{equation}} 
 
\def\gsim{\mathrel{\raisebox{-.6ex}{$\stackrel{\textstyle>}{\sim}$}}} 
\def\sla#1{\ifmmode%
\setbox0=\hbox{$#1$}%
\setbox1=\hbox to\wd0{\hss$/$\hss}\else%
\setbox0=\hbox{#1}%
\setbox1=\hbox to\wd0{\hss/\hss}\fi%
#1\hskip-\wd0\box1 }

\def\asb{{}\ifmmode \bar{\alpha}_s \else $\bar{\alpha}_s$\fi} 
\def \as   {\ifmmode \alpha_s \else $\alpha_s$ \fi}

\def\so3#1{\,{\rm S}_{1,\,3}\left(#1 \right)} 
\def\st2#1{\,{\rm S}_{2,\,2}\left(#1 \right)} 

\newskip\humongous \humongous=0pt plus 1000pt minus 1000pt

\newif\ifdtup

\catcode`@=12 
 
\catcode`\@=11 
 
\@addtoreset{equation}{section} 

\def\theequation{\thesection.\arabic{equation}} 
 
\def\@normalsize{\@setsize\normalsize{15pt}\xiipt\@xiipt 
\abovedisplayskip 14pt plus3pt minus3pt%
\belowdisplayskip \abovedisplayskip 
\abovedisplayshortskip \z@ plus3pt%
\belowdisplayshortskip 7pt plus3.5pt minus0pt} 
 
\def\small{\@setsize\small{13.6pt}\xipt\@xipt 
\abovedisplayskip 13pt plus3pt minus3pt%
\belowdisplayskip \abovedisplayskip 
\abovedisplayshortskip \z@ plus3pt%
\belowdisplayshortskip 7pt plus3.5pt minus0pt 
\def\@listi{\parsep 4.5pt plus 2pt minus 1pt 
     \itemsep \parsep 
     \topsep 9pt plus 3pt minus 3pt}} 
 
\@twosidetrue

\catcode`\@=11  
\def\section{\@startsection{section}{1}{\z@}{3.5ex plus 1ex minus 
   .2ex}{2.3ex plus .2ex}{\large\bf}}

\def\thesection{\arabic{section}} 
\def\thesubsection{\arabic{section}.\arabic{subsection}} 
\def\thesubsubsection{\arabic{section}.\arabic{subsection}.\arabic{subsubsection}} 
 
\def\appendix{\setcounter{section}{0} 
 \def\thesection{\Alph{section}} 
 \def\theequation{\Alph{section}.\arabic{equation}} 
 \def\thesubsection{\Alph{section}.\arabic{subsection}} 
\def\thesubsubsection{\Alph{section}.\arabic{subsection}.\arabic{subsubsection}} 
 
\def\section{\@startsection{section}{1}{\z@}{3.5ex plus 1ex minus 
   .2ex}{2.3ex plus .2ex}{\large\bf}} 
}

\newcommand{\ccaption}[2]{ 
  \begin{center} 
    \parbox{0.85\textwidth}{ 
      \caption[#1]{\small\it {#2}}} 
  \end{center}    } 
 
\def \to   {\mbox{$\rightarrow$}}

\newcount\minutes 
\newcount\scratch 
 
\def\timestamp{%
\scratch=\time 
\divide\scratch by 60 
\edef\hours{\the\scratch} 
\multiply\scratch by 60 
\minutes=\time 
\advance\minutes by -\scratch 
---$\,$\hours:\null 
\ifnum\minutes< 10 0\fi 
\the\minutes}

\begin{document} 
\begin{titlepage} 
\nopagebreak 
{\flushright{ 
        \begin{minipage}{5cm} 
         FTUV-10-1117 \\
         KA-TP-25-2010 \\
         SFB/CPP-10-112 \\
         TTK-10-51
        \end{minipage}        } 
} 
\vfill 
\begin{center} 
{\LARGE \bf \sc 
 \baselineskip 0.9cm 
Gluon-fusion contributions to $\Phi + 2$ Jet production
} 
\vskip 0.5cm  
{\large   
F.~Campanario$^1$,~M.~Kubocz$^2$ and D.~Zeppenfeld$^1$
}   
\vskip .2cm  
$^1${ {\it Institut f\"ur Theoretische Physik, Universit\"at Karlsruhe
    P.O.Box 6980 76128 Karlsruhe, Germany}}\\  
$^2${{\it Institut f\"ur Theoretische Teilchenphysik und Kosmologie, RWTH Aachen
University, D–52056 Aachen, Germany}}\\
\vskip 
1.3cm     
\end{center} 
 
\nopagebreak 
\begin{abstract}
In high energy hadronic collisions a scalar or pseudoscalar Higgs boson,
$\Phi=H$, $A$, can be efficiently produced via gluon fusion, which is 
mediated by heavy quark loops. We here consider double real 
emission corrections to $\Phi=A$ production, which lead to a 
Higgs plus two-jet final state, at order $\alpha_s^4$. Full quark mass effects
are considered in the calculation of scattering amplitudes for the
$\mathcal{CP}$-odd Higgs boson $A$, as induced by quark triangle-, box- and
pentagon-diagrams. They complement the analogous results for a
$\mathcal{CP}$-even Higgs boson $H$ in Ref.~\cite{DelDuca:2001eu}.
Interference effects between loops with top and bottom
quarks as well as between $\mathcal{CP}$-even and $\mathcal{CP}$-odd couplings
of the heavy quarks are fully taken into account. 
\end{abstract} 
\vfill 
\today \timestamp \hfill 
\vfill 
\end{titlepage} 
%



\section{Introduction}
\label{sec:intro}
One of the prime tasks of the CERN Large Hadron Collider (LHC) is the 
search for the origins of the spontaneous breaking of the electroweak
$SU(2)\times U(1)$ gauge symmetry and, once such particles are found,  
the study of one or several Higgs bosons as the remnants of the symmetry 
breaking mechanism. Among the various Higgs boson production channels
the gluon fusion and the weak boson fusion processes have emerged as the most 
promising channels for Higgs boson discovery at the 
LHC~\cite{CMS,ATLAS,Djouadi:2005gj}, and they are equally valuable for
the study of its properties, like the measurement of its couplings  
to gauge boson and fermions~\cite{Zeppenfeld:2000td,Duhrssen:2004cv}.

In weak boson fusion (WBF), $qq\to qqH$ mediated by $t$-channel $W$ or $Z$
exchange, the two forward tagging jets arising from the scattered quarks
provide a tell-tale signature which can be used for efficient background 
rejection~\cite{WBFstudies}. The same $Hjj$ signature can also arise in 
gluon fusion events, via ${\cal O}(\alpha_s^2)$ real emission corrections to
$gg\to H$ which, within the Standard 
Model (SM), is mediated mainly by a top quark loop. For a Higgs boson 
which is lighter than the top quark, the resulting $Hjj$ cross section can
be determined to good approximation by an effective Lagrangian of energy 
dimension five, which is given by~\cite{Kauffman:1996ix,Kauffman:1998yg}
\begin{equation}
{\cal L}_{\rm eff} =
\frac{y_t}{y_t^{SM}}\cdot\frac{\alpha_s}{12\pi v} \cdot H \,
G_{\mu\nu}^a\,G^{a\,\mu\nu} +
\frac{\tilde y_t}{y_t^{SM}}\cdot\frac{\alpha_s}{8\pi v} \cdot A \,
G^{a}_{\mu\nu}\,\tilde{G}^{a\, \mu\nu}\;,
\label{eq:ggS}
\end{equation}
where $G^{a}_{\mu\nu}$ denotes the gluon field strength and
$\tilde{G}^{a\, \mu\nu} = 1/2\, 
G^{a}_{\rho\sigma}\,\varepsilon^{\mu\nu\rho\sigma}$ its dual.
The two terms result from a $\overline{t} tH$ and a $\overline{t}
\ii\gamma_5 tA$
coupling of the (pseudo)scalar Higgs, respectively, and they lead to 
distinctively different distributions of the azimuthal angle between the 
two jets: the $\mathcal{CP}$-even $Hgg$ coupling produces a minimum for 
$\phi_{jj}=\pm\pi/2$ while a $\mathcal{CP}$-odd $Agg$ coupling leads to
minima at $\phi_{jj}=0$ and $\pm\pi$. These distinctions become
important in two-Higgs-doublet models (2HDM) like the minimal
supersymmetric extension of the standard model (MSSM), where a
$\mathcal{CP}$-odd Higgs, $A$, appears in addition to a light and a 
heavy, neutral $\mathcal{CP}$-even Higgs, $h$ and $H$: the azimuthal
angle distribution of $\Phi jj$ events allow to differentiate between a
$\mathcal{CP}$-even Higgs ($\Phi=h,H$) or a $\mathcal{CP}$-odd one
($\Phi=A$). 

For large Higgs boson masses ($m_H\gsim m_t$) the full quark mass 
dependence of the loop diagrams must be calculated for reliable predictions, 
and the same is true for large ratios of the two vacuum expectation values, 
$v_u/v_d=\tan\beta$, where bottom quark loops provide the dominant 
contributions to $qq\to qqH$, $qg\to qgH$ and $gg\to ggH$ amplitudes. 
For a $\mathcal{CP}$-even Higgs boson, these calculations were performed in 
Ref.~\cite{DelDuca:2001eu}. The purpose of the present paper is to present 
the corresponding results for a $\mathcal{CP}$-odd Higgs boson,
$\Phi=A$, or, more precisely for an underlying Higgs coupling to quarks
derived from the Yukawa Lagrangian ${\cal L}=y_q\overline{q} \ii \gamma_5
qA$. By combining the present results with those for a
$\mathcal{CP}$-even Higgs, the quark loop induced contributions to
$\Phi jj$ production can be calculated for an arbitrary Yukawa
coupling of the form
\beq
{\cal L}_{\rm Yukawa}=\overline{q} \, (y_q + \ii \gamma_5
\tilde{y}_q)\, q\,  \Phi \, .
\eeq 
Our results are implemented in a parton level Monte Carlo 
program which is part of the VBFNLO program package~\cite{Arnold:2008rz}. This
numerical implementation allows to calculate $\Phi jj$ production cross
sections in hadronic collisions including top- and bottom-quark loop 
contributions for arbitrary combinations of the Yukawa couplings $y_q$ and
$\tilde{y}_q$ ($q=t,b$).

Our paper is organized as follows. In Section~\ref{sec:calc} we first
define the models in which we consider pseudoscalar Higgs production.
We then provide an
outline of the calculation of  the scattering amplitudes for the three
basic subprocesses, $qq\to qqA$, $qg\to qgA$ and $gg\to ggA$. Further
details on the various loop contributions are relegated to
the Appendices. We have performed a number of analytic and numerical
consistency checks on our calculation: they are described in
Section~\ref{sec:checks}. The main phenomenological results are
presented in Section~\ref{sec:LHCphysics}, for $pp$ scattering at the
LHC with a center of mass energy of $\sqrt{s}=14$~TeV. For various
combinations of top- and bottom quark contributions, parameterized by
$\tan\beta$, we provide integrated $Ajj$ cross sections but also
differential distributions. Results are also presented for general 
$\Phi jj$ events, i.e. for the production of a Higgs boson with arbitrary 
$\mathcal{CP}$-violating couplings to the third generation quarks. Final
conclusions are drawn in Section~\ref{sec:Conclusions}.

\section{Outline of the calculation and matrix elements}
\label{sec:calc}
The production of the $\mathcal{CP}$-odd Higgs boson $A$ in association with
two jets, at order $\alpha_s^4$, proceeds in analogy to the production of  
the $\mathcal{CP}$-even Higgs boson $H_{SM}$ of the SM. The $H_{SM}jj$
production processes with full heavy quark mass effects were considered in 
Ref.~\cite{DelDuca:2001eu}, and we here closely follow the framework and the notation 
introduced there. We consider the production subprocesses
\begin{align}  \label{subproc}
qq \rightarrow qqA \; , \qquad qQ \rightarrow qQA\; , \qquad qg \rightarrow
qgA\; , \qquad gg \rightarrow ggA\; ,
\end{align}
and all crossing-related processes. Here the first two entries 
denote scattering of identical and non-identical quark flavors. 
The Higgs boson $A$ is produced by massive quark loops, for which only 
the third quark generation is taken into account. Furthermore, within
the MSSM, massive squark loops can safely be neglected, because 
their contribution sums to zero at amplitude level in the production of a 
$\mathcal{CP}$-odd Higgs boson~\cite{HHG}. In the 2HDM, up- and down-type
quark Yukawa couplings depend on the ratio of vacuum expectation values,
$\tan \beta = v_u/v_d$, via the relations
\begin{enumerate}
\item \textbf{2HDM of type I}:
\begin{align}
\tilde{y}^{\text{I}}_{A,uu}= \frac{\cot \beta}{v}m_u  
\qquad \text{and} \qquad
\tilde{y}^{\text{I}}_{A,dd}= -\frac{\cot \beta}{v}m_d \; ,
\end{align}

\item \textbf{2HDM of type II (MSSM)}:
\begin{align} \label{c:tb}
\tilde{y}^{\text{II}}_{A,uu}= -\frac{\cot \beta}{v}m_u  
\qquad \text{and}
\qquad
\tilde{y}^{\text{II}}_{A,dd}= -\frac{\tan \beta}{v}m_d  \; .
\end{align}
\end{enumerate}
In the 2HDM of type I, Yukawa couplings for up-type and down-type quarks are
suppressed equally at large $\tan \beta$ compared to the 2HDM of type II, 
where only the up-type Yukawa coupling is suppressed but the down-type 
Yukawa coupling is enhanced. Due to this enhancement, loops with bottom 
quarks can also provide significant contributions to cross sections.

\begin{figure}[thb] 
\centerline{ 
\epsfig{figure=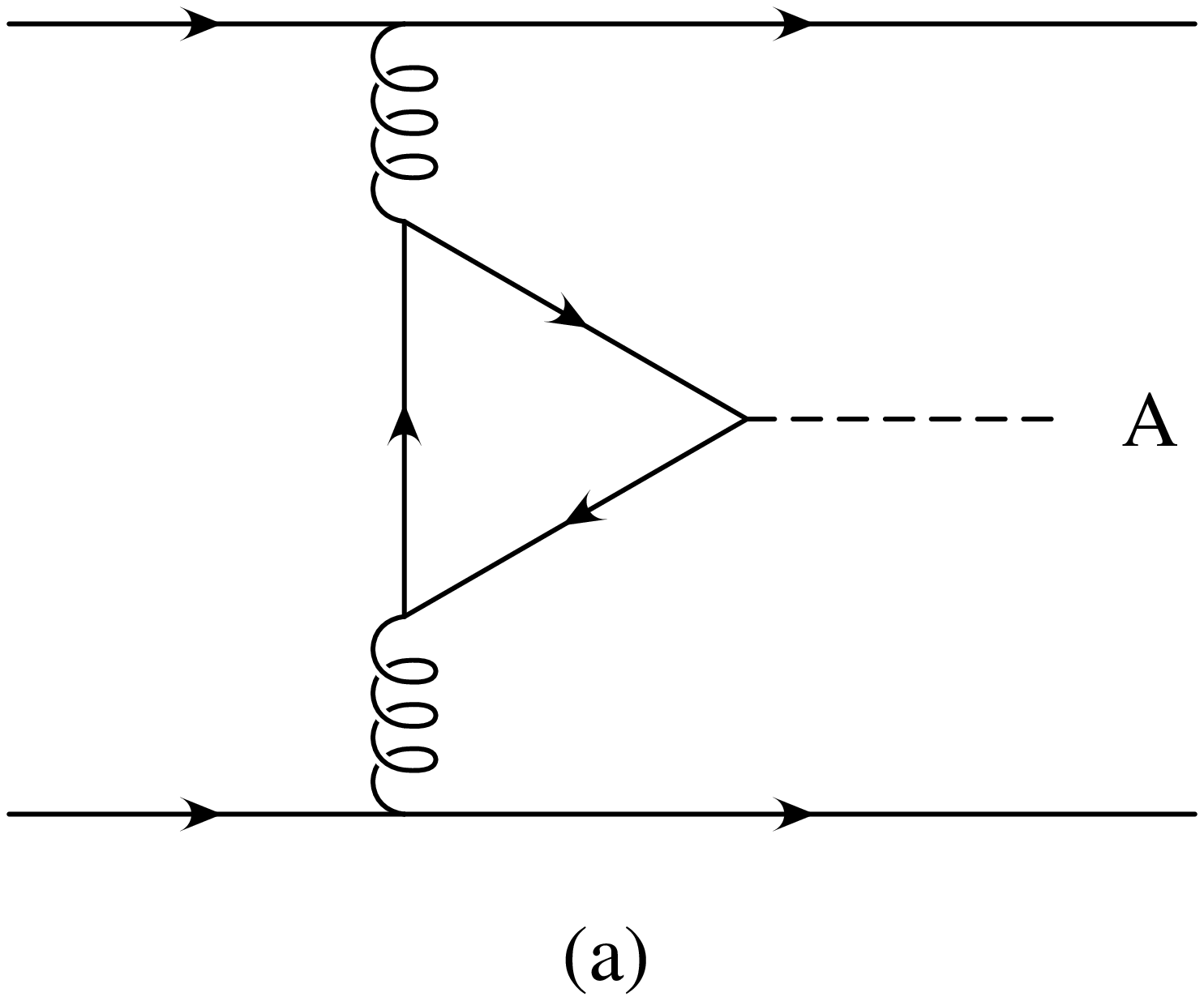,width=4.8cm} \ \  
\epsfig{figure=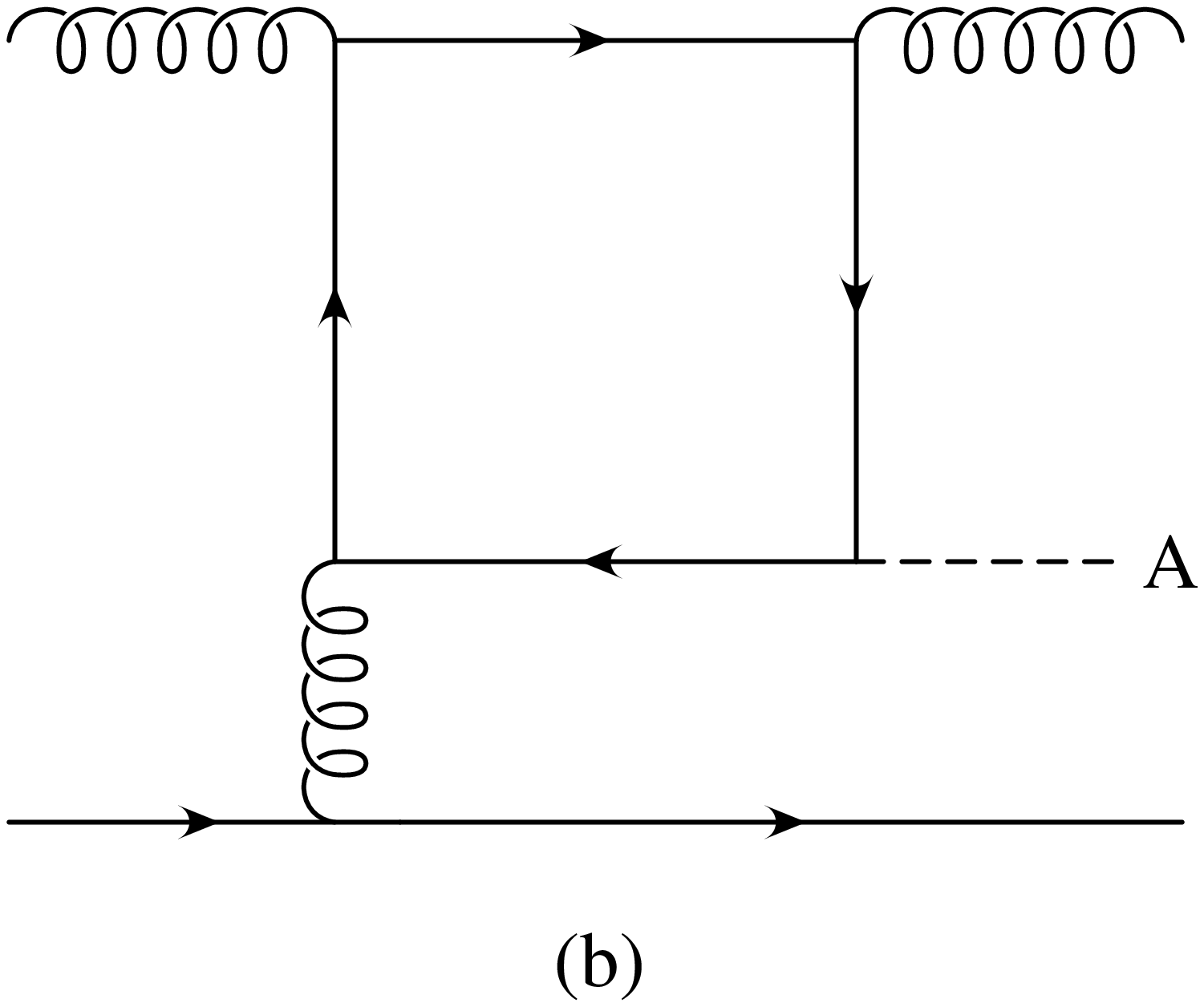,width=4.8cm} \ \  
\epsfig{figure=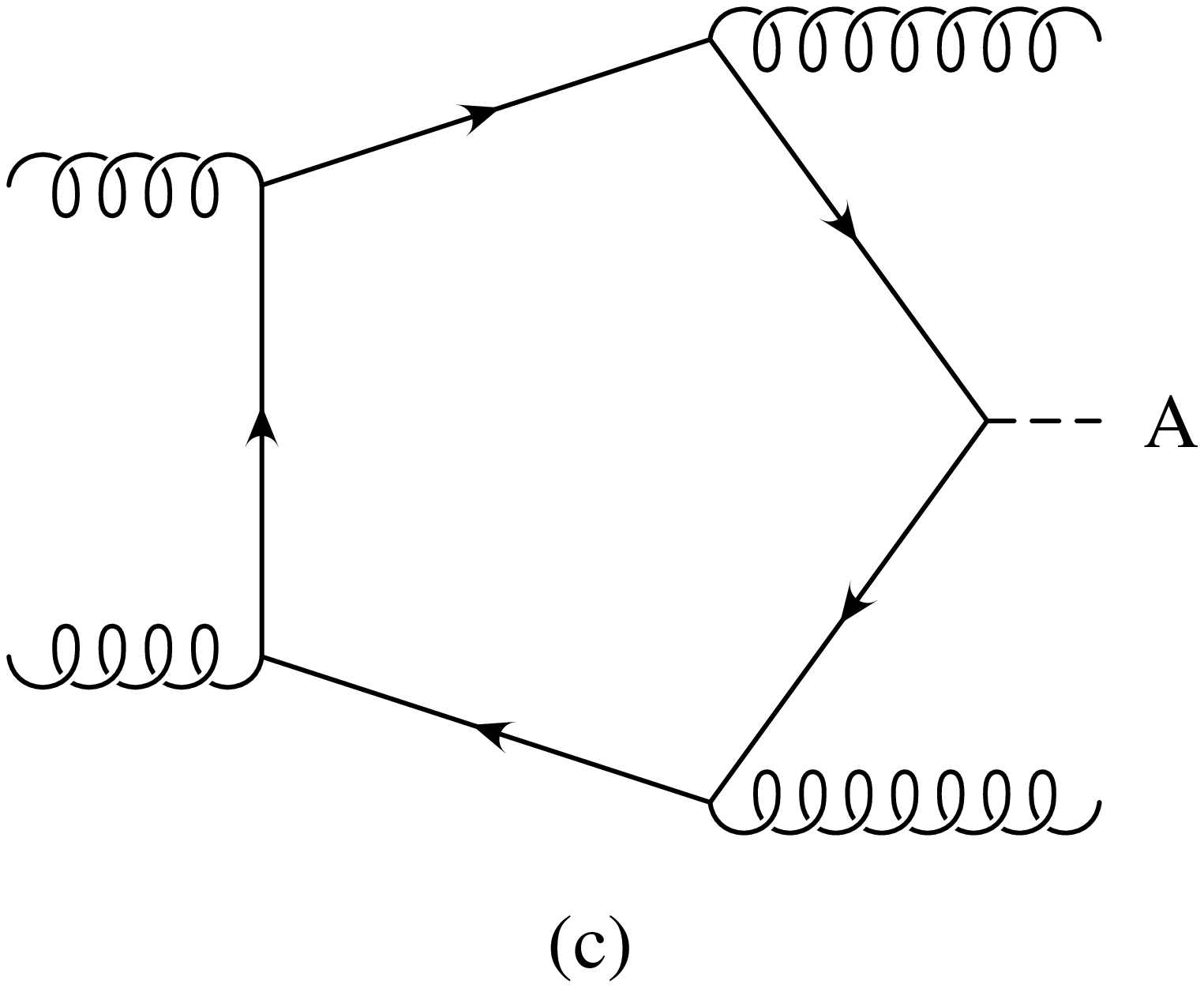,width=4.8cm} \ \  
} 
\ccaption{} 
{ \label{tbp} Examples of Feynman graphs contributing to $A+2$~jet 
production via \newline
\hspace*{1.85cm} gluon fusion. } 
\end{figure}

In the calculation of the subprocesses listed in~\eqref{subproc}, three
different loop topologies appear: the triangle-, box-, and pentagon diagrams 
of Fig.~\ref{tbp}. 
The contributing Feynman diagrams can be easily built from the simpler
QCD dijet processes at leading order. One needs to insert the Higgs-gluon
triangles into the gluon propagators of the $2 \to 2$ tree-level diagrams
in all possible ways or one replaces a triple gluon or 
four gluon vertex by box or pentagon graphs in all 
possible ways. Charge-conjugation related diagrams, where the loop
momentum is running clockwise and counter-clockwise, can be counted as one by
exploiting Furry's theorem~\cite{Furry}. This effectively reduces the number of diagrams 
by a factor of two. Furthermore
all diagrams are UV-convergent and, due to the finite quark mass in the loops,
also IR-convergent.
All coupling constants and loop factors which appear can conveniently be
absorbed into an overall factor
\begin{align} \label{factorF}
F_f=4m_fh_f\frac{g_s^4}{16 \pi^2}=4m_f h_f \alpha_s^2 \; ,
\end{align}
where $f=b,t$ labels the heavy quark flavor of a particular loop.
In the following, we use the MSSM couplings of Eq.~\eqref{c:tb}, i.e. we set
$h_t =\cot \beta \ m_t/v$ and $ h_b=\tan \beta \ m_b /v$. By pulling out a
loop factor $4m_f/16 \pi^2$, we anticipate that the Dirac trace of all
loops requires a quark mass insertion to compensate the helicity flip induced
by the $\Phi f\overline{f}$ coupling. 

\subsection{\boldmath Subprocesses $qQ\to qQA $ and $qq\to qqA$}
The subprocess $qQ\to qQA $, depicted in Fig.~\ref{tbp}(a), is the simplest
contribution to $A+2$ jet production. Following Ref.~\cite{DelDuca:2001eu}, 
the amplitude for different flavors can be written  as
\begin{align}
{\cal A}^{qQ} =\sum_{f=t,b} F^{qQ}_f J_{21}^{\mu_1} J_{43}^{\mu_2}
T_{\mu_1\mu_2}(q_1,q_2,m_f) \, t^a_{i_2i_1}\, t^a_{i_4i_3} = 
{\cal A}_{2143}^{qQ}  \, t^a_{i_2i_1}\, t^a_{i_4i_3} \;. 
\end{align}
Using the notation
and formalism for the spinor algebra of~\cite{Hagiwara:1988pp}, the external
quark lines can be expressed by effective quark currents $J_{21}^{\mu_1}$ and
$J_{43}^{\mu_2}$ as given in~\cite{DelDuca:2001eu}. The triangle tensor
$T^{\mu_1 \mu_2}(q_1,q_2,m_f)$ (see Appendix Fig.~\ref{app:triangle}) has 
the simple form  
\begin{align}
T^{\mu_1 \mu_2}(q_1,q_2,m_f)= \varepsilon^{\mu_1 \mu_2 q_1 q_2}\
C_0(q_1,q_2,m_f) \; .
\end{align}
Here, $C_0$ denotes the scalar three-point function and
$\varepsilon^{\mu_1 \mu_2 q_1 q_2} $ is the totally anti-symmetric tensor 
(Levi-Civita symbol) in four dimensions contracted with attached gluon 
momenta $q_1$ and $q_2$.
The $t^a_{ij}={\lambda^a_{ij}/2}$ are color generators in the fundamental
representation of SU$(N)$, $N=3$ and the overall factor
\begin{align} 
F^{qQ}_f  = S_1\,S_2\,S_3\,S_4\; 
4\sqrt{\overline p_1^0\,\overline p_2^0\,\overline p_3^0\,\overline p_4^0}\;F_f 
\end{align}
includes normalization factors of external quark spinors. Here, the $\overline p_i$
denote physical momenta describing phase space and the wave functions of
fermions and bosons while $p_i$ is used for momenta appearing in the momentum
flow in Feynman diagrams. Both sets of momenta are related by the sign factors
$S_i$~\cite{Hagiwara:1988pp}
\begin{align}
p_i=S_i \ \overline{p}_i \; ,
\end{align}
with $S_i=+1$ for fermions and $S_i=-1$ for anti-fermions. The factor
$F_f$ is given in Eq.~\eqref{factorF}. For identical quark flavors, one has
to keep in mind Pauli interference
\begin{align} 
\mathcal{A}^{qq}=\mathcal{A}^{qq}_{2143}t^a_{i_2 i_1} t^a_{i_4 i_3}
-\mathcal{A}^{qq}_{4123} t^a_{i_4 i_1} t^a_{i_2 i_3} \; .
\end{align}
The squared amplitude, summed over initial- and final-particle color, 
becomes 
\begin{align} 
  \sum_{\text{color}}|\mathcal{A}^{qq}|^2= \Big(|\mathcal{A}_{2143}|^2
+ |\mathcal{A}_{4123}|^2 \Big) \frac{N^2-1}{4} +2 \,
\text{Re}\big(\mathcal{A}_{2143}\mathcal{A}^*_{4123}\big)\frac{N^2-1}{4N}
\; .
\end{align}

\subsection{\boldmath Subprocess $qg\to qgA $}
Polarization vectors  of external gluon lines with a
triangle insertion can be expressed by effective 
polarization vectors
\begin{align}
e_{iA}^{\mu}(m_f)=\varepsilon^{\mu \epsilon_i q_i P} \ \frac{1}{\big( q_i+P
\big)^2} \ C_0 \big( q_i,-(q_i+P),m_f \big) \ ,
\end{align}
which replace the polarization vectors $\epsilon_i^{\mu}$ of the underlying 
$2\to 2$ process for gluons $i=1,2$. Here $q_i$ is the external gluon 
momentum while $P$ denotes the 
momentum of the Higgs boson. The expression for the amplitude of graphs with a
triangle insertion adjacent to a three-gluon vertex differs slightly from that
in~\cite{DelDuca:2001eu} due to the emergence of the Levi-Civita symbol
\begin{align}
\mathcal{A}^{qg}_{\text{tri}}=\sum_f F^{qg}_f & \big[t^{a_1}, t^{a_2}
\big]_{i_1 i_2}\Bigg\{ 2 \Big[ e_{1A} \cdot \epsilon_2 \ J_{21} \cdot q_2  -
e_{1A} \cdot J_{21} \ \epsilon_2 \cdot (p_2-p_1) - e_{1A} \cdot q_2 \ J_{21}
\cdot \epsilon_2  \Big]\nonumber \\
 &  -2 \Big[  e_{2A} \cdot \epsilon_1 \ J_{21} \cdot q_1  -
e_{2A} \cdot q_1 \ J_{21} \cdot \epsilon_1  -  e_{2A} \cdot J_{21} \
\epsilon_1 \cdot (p_2-p_1) \Big]  \nonumber \\
&  +2 \ \varepsilon_{\mu_1 \mu_2 \mu_3 \mu_4} J_{21}^{\mu_2} \Big[  \epsilon_1
\cdot \epsilon_2 \ q^{\mu_3}_1 q^{\mu_1}_2 (p_2-p_1)^{\mu_4} + \big(
\epsilon_2 \cdot q_1 \ \epsilon^{\mu_1}_1 - \epsilon_1 \cdot q_2 \
\epsilon^{\mu_1}_2 \big)  \nonumber \\
& \times  (q_1 + q_2)^{\mu_3} (p_2 - p_1)^{\mu_4} \Big] \
\frac{C_0(p_2 - p_1,q_1 + q_2,m_f)}{(q_1 + q_2)^2} \Bigg\} \; .
\end{align}
Further expressions for amplitudes of graphs with a triangle insertion can be
taken from~\cite{DelDuca:2001eu} replacing $e_{iH}^{\mu}$ by $e_{iA}^{\mu}(m_f)$. The 
tensor structure of the box diagram in Fig.~\ref{tbp}(b) is given by
$\overline{B}_{\mu_1\mu_2\mu_3}(q_1,q_2,q_3,m_f)$, which is shown pictorially in 
 Fig.~\ref{app:box} and given explicitly in 
Appendix~\ref{sec:appB}. Finally the color structure of the $qg \to qg A$
amplitude is given by~\cite{DelDuca:2001eu}
\begin{align}
\mathcal{A}^{qg} = \big( t^{a_1}t^{a_2} \big)_{i_2 i_1} \mathcal{A}^{qg}_{12} +
\big( t^{a_2} t^{a_1} \big)_{i_2 i_1} \mathcal{A}^{qg}_{21} \; \quad
\text{with} \quad \mathcal{A}^{qg}=\sum_f \mathcal{A}^{qg}_f \; .
\end{align}
The indices 12 and 21 label amplitudes with interchanged external
gluons. Thus, the resulting color-summed squared amplitude takes the form
\begin{align}
\sum_{\text{color}}\big|\mathcal{A}^{qg}\big|^2 =\Big(
\big|\mathcal{A}^{qg}_{12}\big|^2 + \big|\mathcal{A}^{qg}_{21}\big|^2
\Big) \frac{\big(N^2 -1 \big)^2}{4N} -2 \text{Re} \Big[
\mathcal{A}^{qg}_{12} \big(\mathcal{A}^{qg}_{21} \big)^* \Big]
\frac{N^2-1}{4N}\; .
\end{align}

\subsection{\boldmath Subprocess $gg\to ggA $}
After inserting suitable loop topologies and application of Furry's theorem,
this process contains 19 graphs with triangle insertions, 18 box contributions
and 12 pentagon diagrams. The pentagon diagrams Fig.~\ref{tbp}(c) enter via
the $P^{\mu_1 \mu_2 \mu_3 \mu_4 }$ tensor (see Appendix
Fig.~\ref{app:pent}~). Full expressions and diagrams can be looked up
in~\cite{DA}. The contributing color structures to the process $gg \to
gg A$ can be expressed by the real-valued color coefficients $c_i$ defined
in~\cite{DelDuca:2001eu}
\begin{align}
c_1  &=\tr \big[ t^{a_1} t^{a_2} t^{a_3} t^{a_4} \big] + \tr \big[ t^{a_1}
t^{a_4} t^{a_3} t^{a_2}  \big] \; ,\nonumber  \\
c_2 & =  \tr \big[ t^{a_1} t^{a_3} t^{a_4} t^{a_2} \big] + \tr \big[ t^{a_1}
t^{a_2} t^{a_4} t^{a_3} \big] \; , \\
c_3 & =  \tr \big[ t^{a_1} t^{a_4} t^{a_2} t^{a_3} \big] + \tr \big[ t^{a_1}
t^{a_3} t^{a_2} t^{a_4}  \big] \; .\nonumber
\end{align}
Evaluation of the color traces yields 
\begin{align} 
c_1  = & \frac{1}{4} \left( \frac{2}{N} \delta^{a_1 a_2} \delta^{a_3 a_4} +
 d^{a_1 a_2 l} d^{a_3 a_4 l} - f^{a_1 a_2 l} f^{a_3 a_4 l}  \right)\; , \nonumber \\
c_2  = & \frac{1}{4} \left( \frac{2}{N} \delta^{a_1 a_3} \delta^{a_4 a_2} +
d^{a_1 a_3 l} d^{a_4 a_2 l} - f^{a_1 a_3 l} f^{a_4 a_2 l}  \right)\; , \label{ci} \\
c_3  = & \frac{1}{4} \left( \frac{2}{N} \delta^{a_1 a_4} \delta^{a_2 a_3} +
d^{a_1 a_4 l} d^{a_2 a_3 l} - f^{a_1 a_4 l} f^{a_2 a_3 l}  \right) \nonumber \; .
\end{align} 
In terms of these color coefficients, the complete amplitude for $gg\to ggA$
can be decomposed into three separately gauge invariant sub-amplitudes 
\begin{align}
\mathcal{A}^{gg}= \sum^3_{i=1} c_i \sum_f \mathcal{A}^{gg}_{i,f} \; .
\end{align}
The sum over colors of the external gluons for the squared amplitude  
becomes 
\begin{align}
\sum_{\text{color}} |{\cal A}^{gg}|^2 = \sum_{i,j=1}^3  {\cal A}^{gg}_{i} 
\({\cal A}^{gg}_{j}\)^* \sum_{\text{color}} c_i \, c_j \; ,
\end{align}
where the color factors are given by
\begin{align}
& \mathcal{C}_1 \equiv \sum_{\text{color}} c_i c_i=\frac{\big( N^2 -1
   \big) \big( N^4 -2N^2 +6 \big)}{8N^2},\ \text{(no sum. over $i$)\; ,} \\
& \mathcal{C}_2 \equiv \sum_{\text{color}} c_i c_j = \frac{\big( N^2 -1
  \big) \big(3- N^2 \big)}{4N^2} , \quad i \neq j \; .
\end{align}
Thus, one finally gets
\begin{align}
\sum_{\text{color}}|\mathcal{A}^{gg}|^2 =\mathcal{C}_1
\sum^3_{i=1} |\mathcal{A}^{gg}_{i}|^2 + \mathcal{C}_2 \sum^3_{i,j=1; \ i
\neq j} \mathcal{A}^{gg}_{i} \big( \mathcal{A}^{gg}_{j} \big)^*\; .
\end{align}

\section{Numerical implementation}
\label{sec:checks}
Analytic expressions for the amplitudes of the previous chapter 
were implemented in the Fortran program {\em
  VBFNLO}~\cite{Arnold:2008rz,VBFNLO}. The tensor reduction of the
loop contributions up to boxes is performed via Passarino-Veltman
reduction~\cite{Passarino:1978jh}. Additionally we avoid the explicit
calculation of the inverse of the Gram matrix by solving system of
linear equations, which is numerically more stable close to the singular
points. For pentagons, we use the
Denner-Dittmaier algorithm~\cite{Denner:2002ii} which avoids the
inversion of small Gram determinants, in particular for planar
configurations of the Higgs and the two final state partons. 
The program was numerically tested in several ways. Besides usual
gauge-invariance and 
Lorentz-invariance tests of the amplitudes, the different topologies 
were checked separately. The contraction of a triangle-tensor 
$T_{\mu_1 \mu_2}(q_1,q_2,m_f)$ with gluon momentum $q_i^{\mu}$ has to
vanish due to total antisymmetry of the Levi-Civita symbol
\begin{align}
q^{\mu_1}_1 T_{\mu_1 \mu_2} (q_1,q_2,m_f) =q^{\mu_2}_2 T_{\mu_1 \mu_2}
(q_1,q_2,m_f) =0 \; .
\end{align}
Contracting with external gluon momenta, the
tensor expressions of boxes and pentagons reduce to differences of
triangles and boxes, respectively. With the tensor integrals as defined 
in the Appendix, the Ward identities for the boxes read
\begin{align} 
q^{\mu_1}_1 B_{\mu_1 \mu_2 \mu_3}(q_1,q_2,q_3,m_f) &= T_{\mu_2
\mu_3} (q_{12},q_3,m_f) - T_{\mu_2 \mu_3}(q_2,q_3,m_f) \; , \\
q_2^{\mu_2} B_{\mu_1 \mu_2 \mu_3}(q_1,q_2,q_3,m_f) &= T_{\mu_1
\mu_3}(q_1,q_{23},m_f) -T_{\mu_1 \mu_3}(q_{12},q_3,m_f) \; , \\
q_3^{\mu_3} B_{\mu_1 \mu_2 \mu_3}(q_1,q_2,q_3,m_f) &= T_{\mu_1
\mu_2}(q_1,q_2,m_f) -T_{\mu_1 \mu_2}(q_1,q_{23},m_f) \; ,
\end{align}
where the abbreviation $q_{ij}=q_i+q_j$ has been used. Similarly, for the
pentagons one finds 
\begin{align} 
q^{\mu_1}_1 P_{\mu_1 \mu_2 \mu_3 \mu_4}(q_1,q_2,q_3,q_4,m_f)&=B_{\mu_2 \mu_3
\mu_4}(q_{12},q_3,q_4,m_f) -B_{\mu_2 \mu_3 \mu_4}(q_2,q_3,q_4,m_f) \; , \\
q^{\mu_2}_2 P_{\mu_1 \mu_2 \mu_3 \mu_4}(q_1,q_2,q_3,q_4,m_f)&=B_{\mu_1 \mu_3
\mu_4}(q_1,q_{23},q_4,m_f) -B_{\mu_1 \mu_3 \mu_4}(q_{12},q_3,q_4,m_f) \; ,\\
q^{\mu_3}_3 P_{\mu_1 \mu_2 \mu_3 \mu_4}(q_1,q_2,q_3,q_4,m_f)&=B_{\mu_1 \mu_2
\mu_4}(q_1,q_2,q_{34},m_f) -B_{\mu_1 \mu_2 \mu_4}(q_1,q_{23},q_4,m_f) \; , \\
q^{\mu_4}_4 P_{\mu_1 \mu_2 \mu_3 \mu_4}(q_1,q_2,q_3,q_4,m_f)&=B_{\mu_1 \mu_2
\mu_3}(q_1,q_2,q_3,m_f) -B_{\mu_1 \mu_2 \mu_3}(q_1,q_2,q_{34},m_f) \; .
\end{align}
These relationships were tested numerically and they, typically, are 
satisfied at the $10^{-9}$ level when using Denner-Dittmaier reduction 
for the tensor integrals. In addition, one can perform a QED-check for
the pentagons. 
Replacing gluons by photons and considering the process
$\gamma \gamma \to \gamma \gamma A$, diagrams with three- and
four-gluon-vertices vanish, because these structures are not available in an
Abelian theory. The amplitude is simply 
given by the sum of all pentagon graphs,
without color factors. When contracting with an external gauge boson 
momentum one obtains zero, since boxes are not allowed for photons,
by Furry's theorem. Our amplitudes pass this test as well.

To check the full scattering amplitudes, one can make use of the 
heavy-top effective Lagrangian for a SM strength Yukawa coupling,
\begin{align} \label{effL}
\mathcal{L}_{\text{eff}}^A =\frac{\alpha_s}{8 \pi v} G^a_{\mu_1 \mu_2}
\tilde{G}^{a \mu_1 \mu_2} A \quad \text{with} \quad \tilde{G}^{a \mu_1 \mu_2}
= \frac{1}{2} \epsilon^{\mu_1 \mu_2 \mu_3 \mu_4} G^a_{\mu_3 \mu_4}\; .
\end{align}
As $m_t$ becomes large, the results calculated with full fermion loops 
must approach the approximate ones derived from the effective Lagrangian.
This check was performed with $m_t=5000$ GeV, and cross sections
agreed well for Higgs boson masses in the range
$100$ GeV$<m_A<$ 1000 GeV. 
In production runs, we
put a cut in the routines for the determination of the tensor integral
coefficients of the C and D functions such that the complete amplitude
is set to zero when small Gram determinants appear. We have checked that
the result and plots do not depend on this cut for a broad range of values.
Finally, the amplitudes for all three subprocesses were recalculated
using the {\em FeynCalc/FormCalc} framework~\cite{FA}. For a selection of
randomly generated phase space points, the two independent calculations
yield agreement at least at the $10^{-6}$ level.

\section{Applications to LHC physics}
\label{sec:LHCphysics}
The numerical analysis of the $\Phi+2$ jet cross section was performed with 
a parton level Monte Carlo program in the {\em VBFNLO} 
framework~\cite{VBFNLO}, using the CTEQ6L1~\cite{CTEQ6} set for 
parton-distribution functions. 
In order to prevent soft or collinear divergencies in the cross
sections, a minimal set of acceptance cuts has to be
introduced. Following Ref.~\cite{DelDuca:2001eu}, we impose
\begin{align} \label{ICuts}
p_{Tj} > 20 \ \text{GeV}\;, \qquad |\eta_j| < 4.5\;, \qquad R_{jj} > 0.6 \; ,
\end{align}
where $p_{Tj}$ is the transverse momentum of a final state parton and $R_{jj}$
describes the separation of the two partons in the pseudo-rapidity 
versus azimuthal-angle plane
\begin{align}
R_{jj} = \sqrt{\Delta \eta_{jj}^2 + \phi_{jj}^2}\ \;,
\end{align}
with $\Delta \eta_{jj} = |\eta_{j1}-\eta_{j2}|$ and 
$\phi_{jj} = \phi_{j1}-\phi_{j2}$. 
These cuts anticipate LHC detector capabilities and jet finding
algorithms and will be called ``inclusive cuts'' (IC) in the
following. Unless specified otherwise, the factorization scale is set to
\begin{align}
\mu_f =\sqrt{p_{T1} \ p_{T2}}
\end{align}
while the renormalization scale is fixed by setting~\cite{DelDuca:2001eu} 
\begin{align}
\alpha_s^4 (\mu_R) = \alpha_s(p_{T1}) \alpha_s (p_{T2}) \alpha_s^2(m_A) \;.
\end{align}
We use one-loop $\alpha_s$ running with $\alpha_s(M_Z)=0.13$. 
All our results below contain the contributions from the full 
top- and bottom-quark loops. For the
top-quark mass we use $m_t=173.1$~GeV. In the case of
bottom-loops, running Yukawa coupling and propagator mass are taken into account,
with the Higgs-mass as reference scale. Within the
Higgs-mass range of 100 to 600 GeV, the bottom-quark mass is 33 to 42 \% smaller than
the pole mass of $4.855$ GeV. The evolution of $m_b$ up to a reference
scale $\mu$ can be expressed as
\begin{align}
\overline{m}_b (\mu)= \overline{m}_b \left(m_b\right)
\frac{c\big[\alpha_s(\mu) /\pi \big]}{c\big[\alpha_s(m_b) / \pi \big]}
\, ,
\end{align}
with $\overline{m}_b \left(m_b\right)=4.2$ GeV, as derived from the 
relation between pole mass and MS-bar mass.
For the coefficient function $c$, the five flavor
approximation~\cite{Spira:1997dg,Vermaseren:1997fq} within the mass
range $m_b < \mu < m_t$,
\begin{align}
c(x)=\left(\frac{23}{6} \, x \right)^{\frac{12}{23}} \big[1+1.17549 \,
x+1.50071 \, x^2 +0.172478 \, x^3 \big]~,
\end{align}
is used. Further evolution of $\overline{m}_b$ to a
renormalization scale $\mu>m_t$ can be performed safely within the
five flavor approximation, because the deviation to the six
flavor scheme is less than 1\% for $\mu<600$~GeV.
\begin{figure}[t!]
\centering
\includegraphics[scale=1.2]{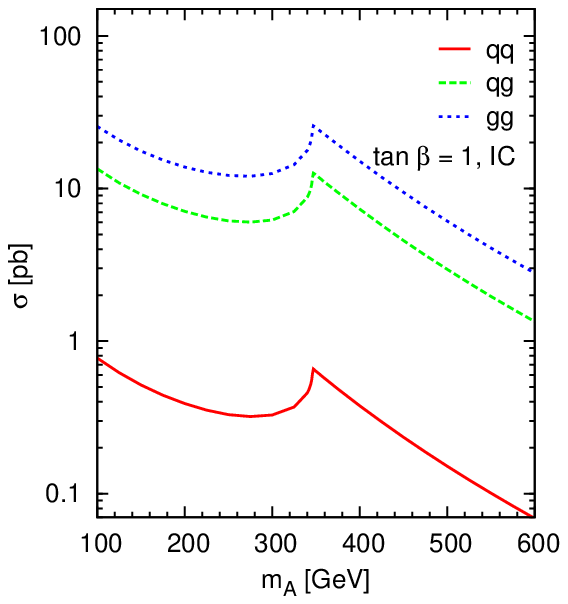}
\ccaption{} 
{\label{mAscanIC}
$A+2$ jet cross section of the individual contributions of the
subprocesses quark-quark, quark-gluon and gluon-gluon scattering for $\tan
\beta=1$ as a function of the pseudo-scalar Higgs boson mass,
$m_A$. Here, the inclusive cuts (IC) of Eq.~\eqref{ICuts} were
applied.}
\end{figure}

\begin{figure}[t!]
\centering
\includegraphics[scale=1.2]{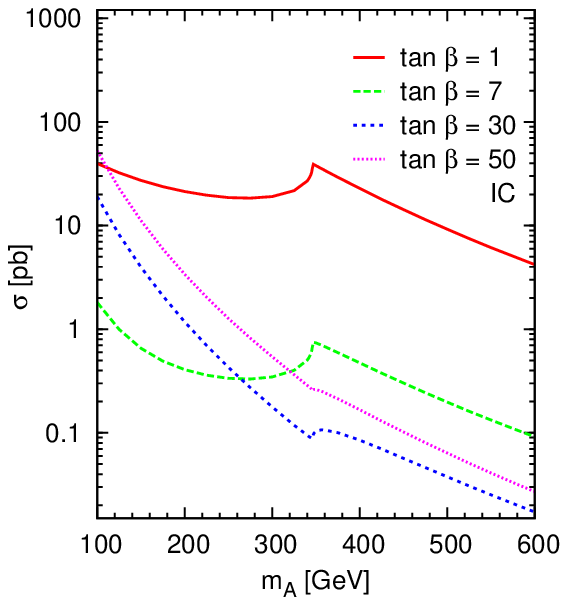}
\includegraphics[scale=1.2]{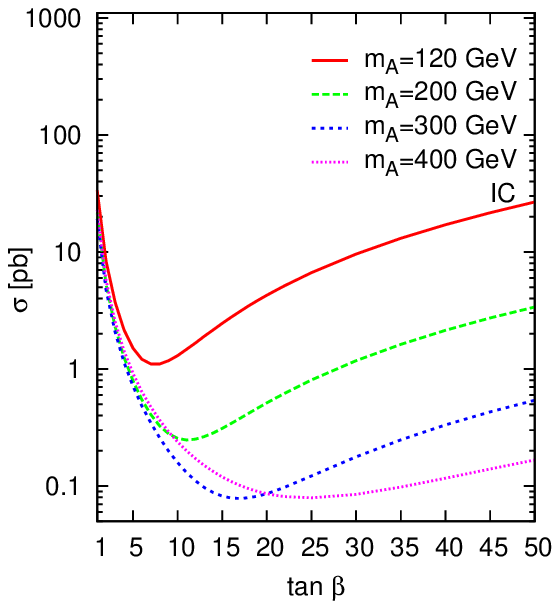}
\ccaption{} 
{\label{tanBscan} 
$A+2$ jet cross section as a function of the pseudo-scalar Higgs boson
mass, $m_A$ for different values of $\tan \beta$. The inclusive cuts
(IC) of Eq.~\eqref{ICuts} are applied}

\end{figure}
Contributions of individual subprocess categories to the total cross
section for $\tan \beta=1$ are shown in Fig.~\ref{mAscanIC} as a
function of the Higgs boson mass, $m_A$. Here, the minimal cuts of
Eq.~\eqref{ICuts} were used. The cross sections for processes involving
gluons (quark-gluon or gluon-gluon amplitudes) exceed the quark-quark scattering contributions by more than one
order of magnitude. 
The $m_A$ dependence of the full cross section, with top- and bottom-quark
interference, is given in the left panel of Fig.~\ref{tanBscan} 
for a selection of $\tan \beta$ values. For small
$\tan \beta$, amplitudes with a top-quark loop dominate over 
bottom-quark loop mediated contributions. The striking peak arises
due to threshold enhancement near $m_A \approx 2 \ m_t$, 
whereas for bottom-quark loop dominated processes the peak would appear well
below the Higgs mass range shown. 


For low $m_A$, the minimal cross section is
obtained near $\tan \beta \approx 7$, when $h_t \approx h_b$ (see
Eq.~\eqref{c:tb}) and both Yukawa couplings are suppressed compared to
$h_t^{SM}$. For large $\tan \beta$, e.g. $\tan \beta=50$ in
Fig.~\ref{tanBscan}(a), the bottom-quark loops dominate. However, they lead
to a much more rapid decrease of the cross section with rising $m_A$
because the suppression scale of the loops
is now set by the heavy Higgs boson mass instead of the quark mass.
The reduced importance of the bottom-quark loops at large $m_A$ 
implies that equality of the top and bottom contributions and, thereby,
the minimum of the production cross section is reached at increasingly
larger $\tan\beta$ as $m_A$ is increased. This effect is demonstrated in
the right panel of Fig.~\ref{tanBscan}. 

\begin{figure}[h] 
\centering
\includegraphics[scale=1.2]{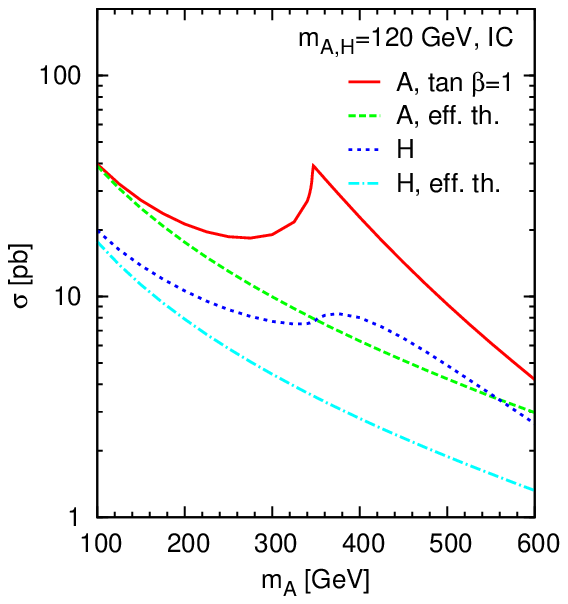}
\ccaption{} 
{\label{effHA} 
Comparison of cross sections for the $\mathcal{CP}$-odd and
$\mathcal{CP}$-even Higgs for the full loop calculation and within the
effective theory. The inclusive cuts (IC) of
Eq.~\eqref{ICuts} are applied.}
\end{figure}

The $\gamma_5$-matrix in the Dirac-trace of the quark loops
leads to a new tensor structure and to a normalization of the loops 
that, for equal Yukawa couplings, induces a $(3/2)^2=2.25$ times
larger $Ajj$ than $Hjj$ cross section. This enhancement is shown in
Fig.~\ref{effHA} and is also apparent in the
effective Lagrangian of Eq.~(\ref{eq:ggS}), where the 
coefficient of the $\mathcal{CP}$-odd $Agg$ coupling exceeds that of the
$Hgg$ coupling by a factor $3/2$. This effective Lagrangian provides a
good approximation to the total $\Phi jj$ cross sections up to
Higgs-masses of $\approx 160$~GeV  and for small transverse
momenta, $p_{Tj}\lesssim m_t$. In this region, the effective Lagrangian 
approximation can be used as a numerically fast alternative for
phenomenological studies~\cite{DelDuca:2001eu}. 

\begin{figure}[ht] 
\centering
\includegraphics[scale=0.98]{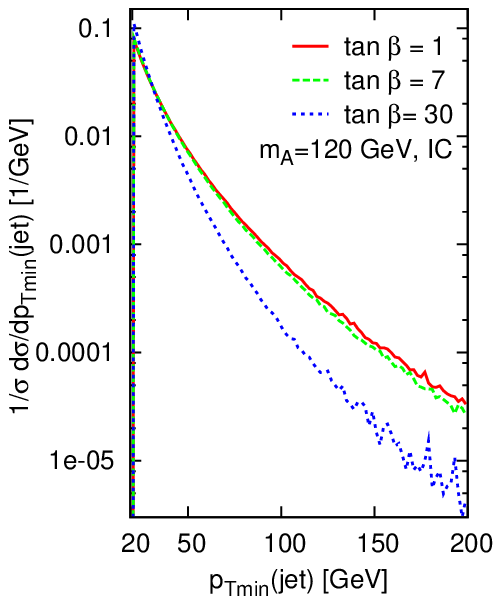}
\includegraphics[scale=0.98]{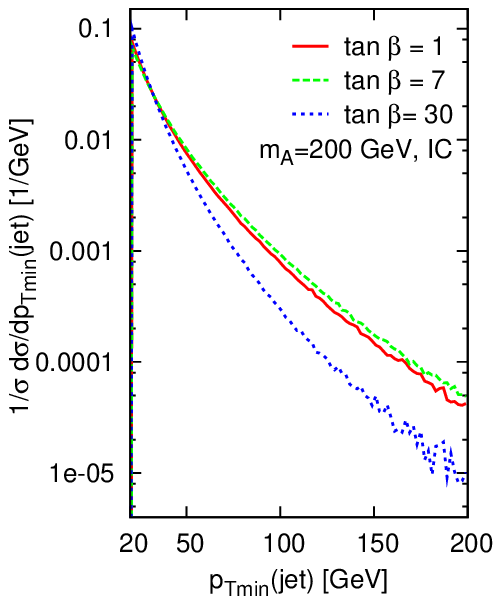}
\includegraphics[scale=0.98]{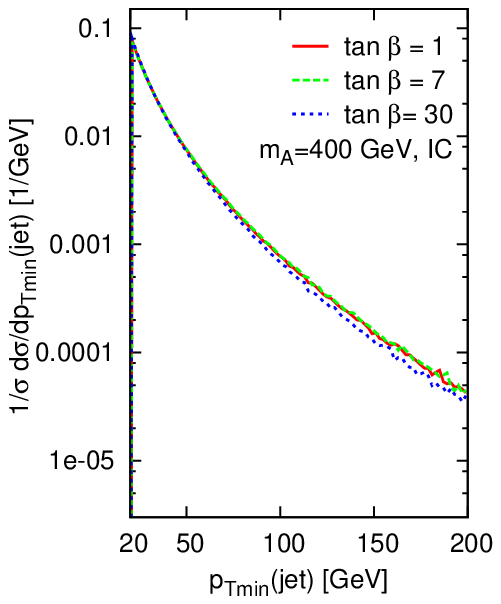}
\ccaption{} 
{\label{fig:pTmin}
Normalized transverse-momentum distributions of the softer
jet in $Ajj$ production at the LHC, for different $\tan \beta$ 
and Higgs-mass values. The inclusive selection cuts of 
Eq.~\eqref{ICuts} are applied.}
\end{figure}

\begin{figure}[ht!] 
\centering
\includegraphics[scale=0.98]{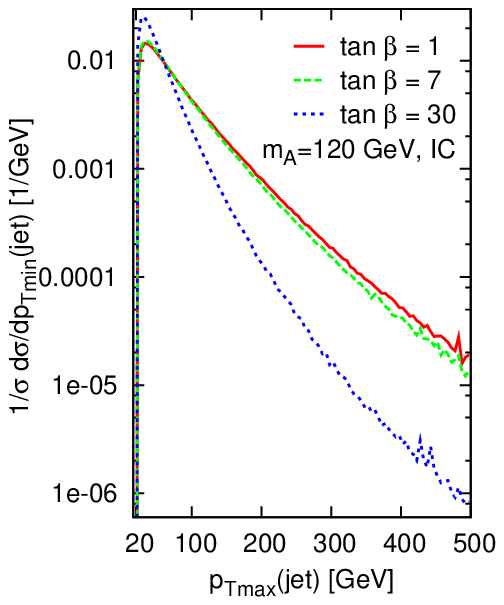}
\includegraphics[scale=0.98]{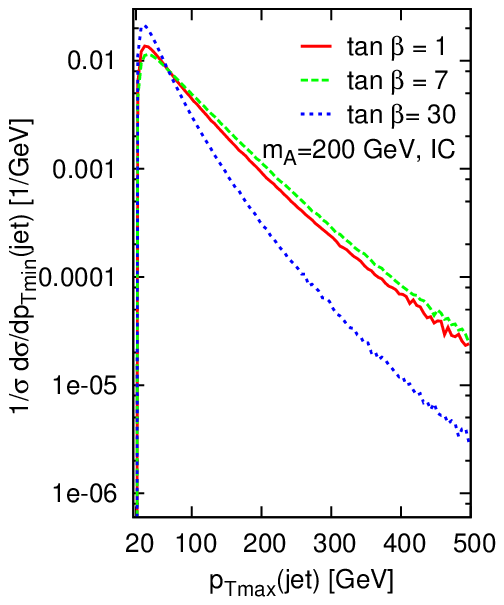}
\includegraphics[scale=0.98]{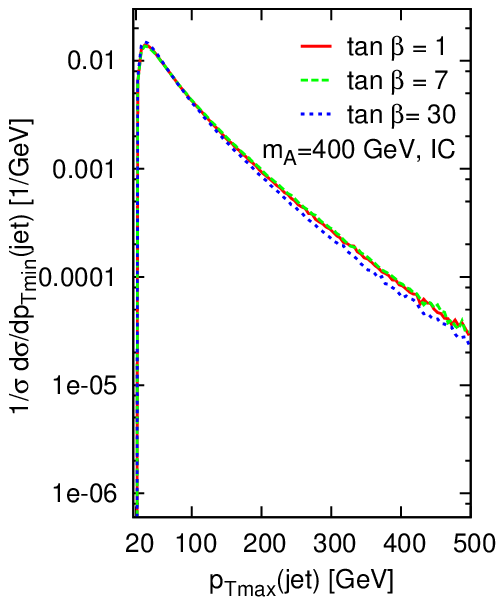}
\ccaption{} 
{\label{fig:pTmax}
Normalized transverse-momentum distributions of the harder
jet in $Ajj$ production at the LHC for different $\tan \beta$ 
and Higgs-mass values. The inclusive selection cuts of 
Eq.~\eqref{ICuts} are applied.}
\end{figure}
The smaller quark mass in the bottom loops also has a pronounced effect 
on the transverse momentum distribution of the accompanying jets:
for $p_{Tj}\gsim m_b$ the large scale of the kinematic invariants leads to
an additional suppression of the bottom induced sub-amplitudes compared 
to the heavy quark effective theory. This effect is clearly visible
in Figs.~\ref{fig:pTmin} and~\ref{fig:pTmax}, where the 
transverse-momentum distributions of the softer and the harder of the
two jets are shown for pseudoscalar Higgs masses $m_A=120$, $200$ and 
$400$~GeV for $\tan\beta=1,\, 7,\, 30$. For modest Higgs mass values, 
both distributions fall more steeply for large $\tan\beta$. At large values
of $m_A$, the Higgs boson mass sets the scale for the fermion loops which, 
in the $m_A=400$~GeV panels of Figs.~\ref{fig:pTmin} and~\ref{fig:pTmax},
leads to $p_T$ distributions which are approximately equal for the top- 
or bottom-quark dominated loops.

\begin{figure}[t] 
\centering
\includegraphics[scale=1.2]{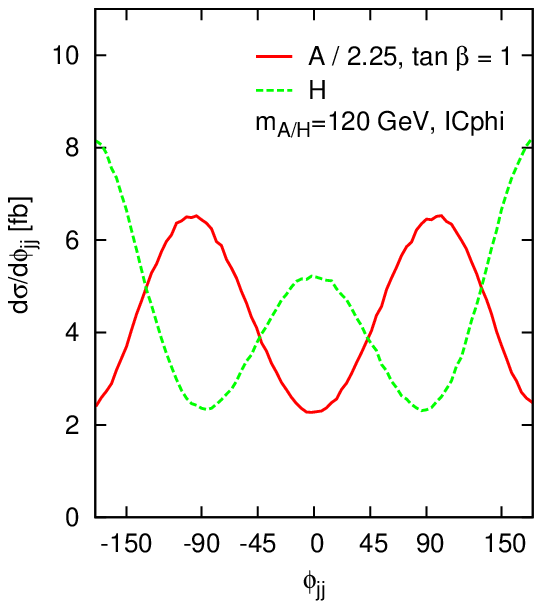}
\includegraphics[scale=1.2]{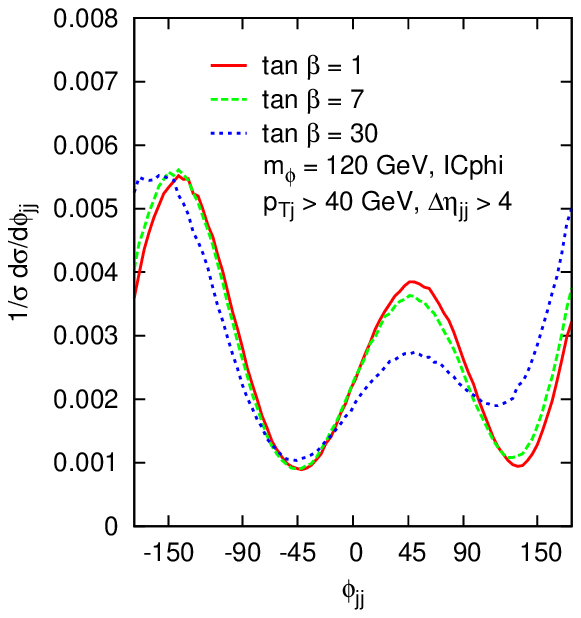} 
\ccaption{} 
{\label{ICphi2} $\phi_{jj}$ distributions for different Higgs sector scenarios: 
pure $\mathcal{CP}$-odd or $\mathcal{CP}$-even coupling in the effective 
Lagrangian limit (left panel) and the $\mathcal{CP}$-violating case 
defined in Eq.~\ref{eq:cpmix} (right panel). In the effective Lagrangian, the
$\mathcal{CP}$-odd coupling was matched to the $\mathcal{CP}$-even
coupling by a factor $2/3$. For the $\mathcal{CP}$-violating Higgs-sector,
results are shown for different $\tan \beta$ values and with jet acceptance cuts tightened to
$\Delta\eta_{jj}>4$ and $p_{Tj}>40$~GeV as compared to the ICphi set~\eqref{ICphi}.} 
\end{figure}

The azimuthal angle between the more forward and the more backward of the 
two tagging jets, $\phi_{jj}=\phi_{jF} -\phi_{jB}$, provides a sensitive 
probe for the $\mathcal{CP}$-character of the Higgs couplings to the 
quarks~\cite{Plehn:2001nj,Hankele:2006ma,Klamke:2007cu}. As shown in 
the left panel of Fig.~\ref{ICphi2} for a heavy quark in the loop, the 
$\mathcal{CP}$-even $Hqq$ coupling produces a minimum for 
$\phi_{jj}=\pm 90$ degrees while a $\gamma_5$-induced $\mathcal{CP}$-odd 
$Aqq$ coupling leads to minima at $\phi_{jj}=0$ and $\pm 180$
degrees. The softening  
effects observed for the jet transverse momentum distribution then raise
the question, to what extent the jet azimuthal angle correlations 
of the effective theory will get modified when bottom quark loops dominate.
 
\begin{figure}[htb] 
\centering
\includegraphics[scale=0.98]{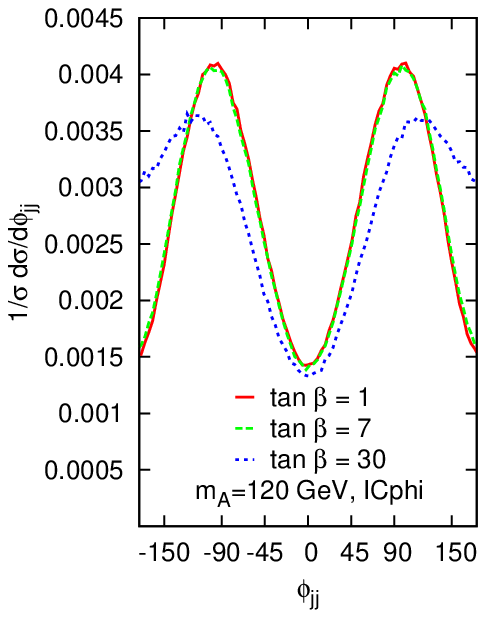}
\includegraphics[scale=0.98]{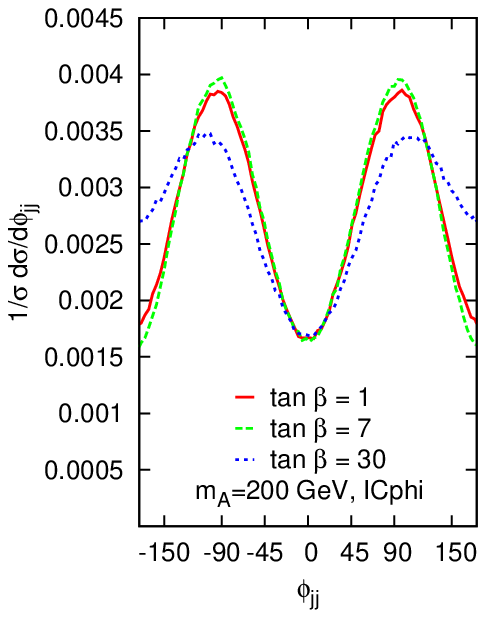}
\includegraphics[scale=0.98]{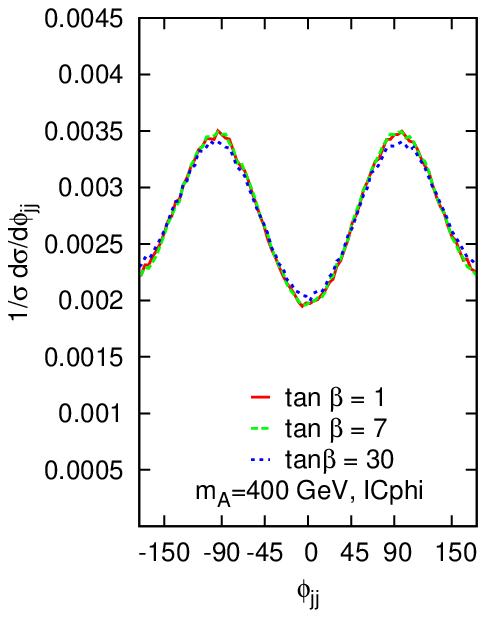}
\ccaption{} 
{\label{ICphi3} Distribution in the azimuthal-angle between the two 
final state jets for $\mathcal{CP}$-odd Higgs boson production at 
different Higgs-masses and $\tan \beta$ values. The ICphi set of acceptance
cuts (see Eq.~(\ref{ICphi})) is used for $pp$ collisions at 14~TeV.} 
\end{figure}

For the set of pseudoscalar Higgs masses and $\tan\beta$ values mentioned
above, predictions for the normalized $\phi_{jj}$-distributions are
shown in Fig.~\ref{ICphi3}. The calculation was carried
out with a modified set of cuts, however, which was shown in 
Ref.~\cite{Klamke:2007pn} to lead to a better sensitivity to the 
$\mathcal{CP}$-structure of the Higgs couplings than the inclusive cuts. 
In  Fig.~\ref{ICphi3}, we use
\begin{align} \label{ICphi}
p_{Tj} > 30 \ \text{GeV}\;, \qquad |\eta_j| < 4.5\;, \qquad R_{jj} > 0.6\;, \qquad
\Delta \eta_{jj}=|\eta_{j1} -\eta_{j2}|>3 \;,
\end{align}
which we call ICphi cuts in the following. One finds that the characteristic
structure of the $\phi_{jj}$ distribution, dips at $\phi_{jj}=0$ and
$\phi_{jj}=\pm 180$ degrees, 
remains for bottom quark dominated $Ajj$ production, 
albeit at a quantitatively reduced level for $m_A> 2m_q$. For a relatively
light pseudoscalar Higgs boson and large $\tan\beta$, the softer 
transverse momentum distribution of the Higgs leads to kinematical
distortions of the $\phi_{jj}$ distribution: at $\phi_{jj}\approx 0$
the Higgs recoils against two jets and hence must have $p_{TH}>60$~GeV, 
and this high $p_T$-scale leads to an additional suppression as compared 
to the $\phi_{jj}\approx \pm 180$ degree case where transverse momentum
balancing of the jets does allow $p_{TH}=0$. 

For the effective theory of the large quark mass limit, it was observed that 
$\mathcal{CP}$-violating effects due a mixture of $\mathcal{CP}$-even and 
$\mathcal{CP}$-odd couplings leads to a phase shift of the $\phi_{jj}$ 
distribution compared to the $\mathcal{CP}$-even case by an angle, $\alpha$,
which is given by the relative strength of the two 
couplings~\cite{Hankele:2006ma,Klamke:2007cu}. Taking into 
account the relative enhancement by the factor $3/2$ of the $Agg$ coupling
due to loop effects, the phase shift angle is given by
\beqn
\tan\alpha = \frac{3}{2}\, \frac{\tilde{y}_q}{y_q}
\eeqn
when heavy quark loops of a single flavor dominate. In order to test this
effect for the case of a light quark, we show, in the right panel of 
Fig.~\ref{ICphi2}, the results for 
\begin{align}
\label{eq:cpmix}
y_b = \frac{3}{2}\tilde{y}_b=-\tan\beta \frac{m_b}{v}
\qquad {\rm and} \qquad
y_t = \frac{3}{2}\tilde{y}_t=-\cot\beta \frac{m_t}{v}\; ,
\end{align}
where a 45 degree phase shift is expected, with minima of the
$\phi_{jj}$ distribution at -45 and +135 degrees. This basic expectation
is, indeed, confirmed by the detailed calculation. However, there are
additional distortions of the azimuthal angle distributions which can
again be explained by kinematical effects due to transverse momentum
balancing of the two jets and the Higgs boson.

\section{Conclusions}
\label{sec:Conclusions}

In this paper, we have presented the determination of quark mass effects
on the cross section and on distributions for pseudoscalar Higgs
production in association with two final state partons. Our calculation
for $Ajj$ production complements the analogous one for a scalar Higgs,
i.e. $Hjj$ production as carried out in 
Ref.~\cite{DelDuca:2001eu}. Qualitative features are quite similar for
the two cases. Validity of the heavy quark mass approximation is found
to be restricted to $m_\Phi< m_q$ and $p_{Tj}<m_q$ while large dijet
invariant 
masses do not spoil the validity of the heavy quark limit. A pronounced
difference between $Ajj$ and $Hjj$ production is observed in the
azimuthal angle distribution between the two jets, which allows, in
principle, to determine the $\mathcal{CP}$-properties of the produced
Higgs boson at the LHC~\cite{Klamke:2007cu}.

Our analytical expressions have been implemented in the VBFNLO
program~\cite{Arnold:2008rz} and are publicly available as a parton
level Monte Carlo program. Even though the code must evaluate loop
expressions up to pentagons, the calculation is leading order in the
strong coupling constant since Higgs production in gluon fusion first
appears at the one-loop level. As a leading order process, it has been
provided with an interface in the Les Houches format~\cite{Boos:2001cv} 
to run with parton shower programs, providing full particle, momentum and
color flow information. The code allows to sum top- and bottom-quark
induced contributions with arbitrary $\mathcal{CP}$-violating couplings
\beq
{\cal L}_{\rm Yukawa}=\overline q (y_q + \ii \gamma_5 \tilde{y}_q)q \Phi  \,,
\eeq
and, thus, is versatile enough for simulating the effects of 
general, $\mathcal{CP}$-violating Higgs sectors at the LHC.

\appendix
\section{Tensor structure of triangles}
\label{sec:appA}
\begin{figure}[ht] 
\centering
\includegraphics[scale=0.35,angle=0]{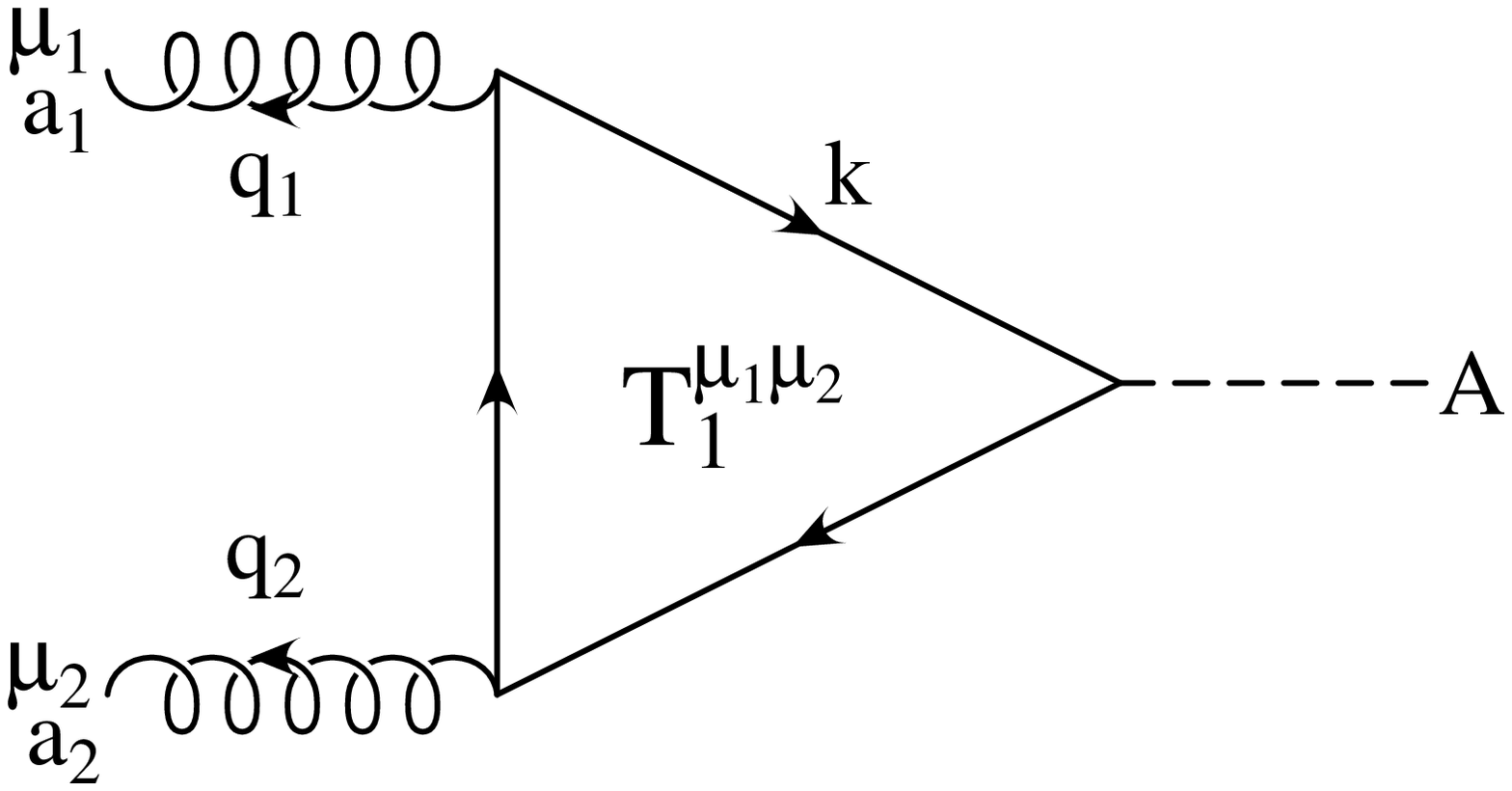} \hspace{1cm}
\includegraphics[scale=0.35,angle=0]{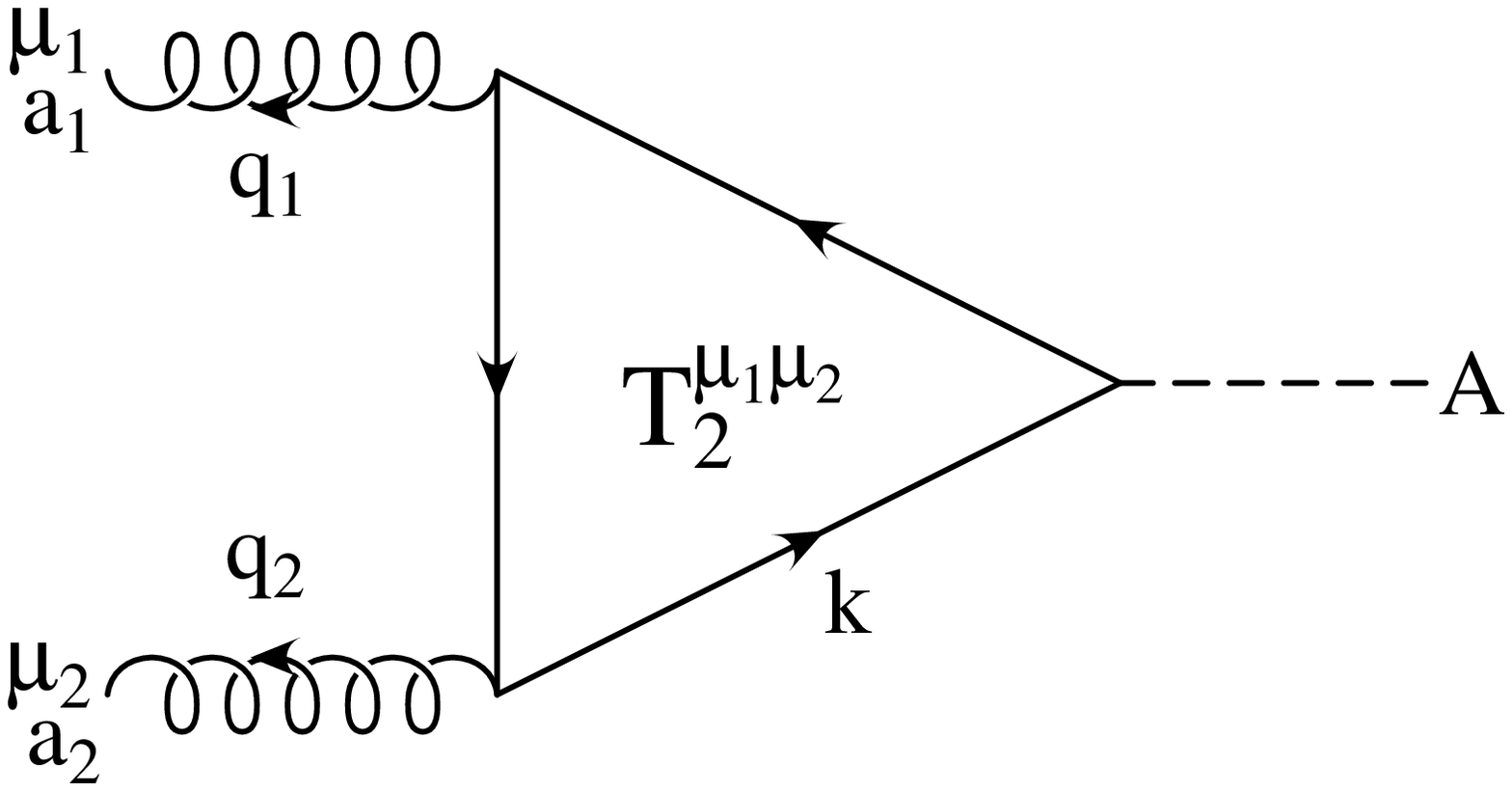}
\ccaption{}
{ \label{app:triangle} Two three-point functions connected by charge 
               conjugation.}  
\end{figure} 

The generic three-point functions for triangle graphs with opposite loop
momentum have the following expressions

\begin{align}
T^{\mu_1 \mu_2}_1(q_1,q_2,m_f)= \frac{-\ii}{4m_f} \int \frac{\id^4k}{\ii\pi^2}
\tr \Bigg[ & \frac{\fdag{k}+ m_f}{k^2-m_f^2} \gamma^{\mu_1} \frac{\fdag{k}+
\fdag{q_1}+m_f}{(k+q_1)^2-m_f^2} \gamma^{\mu_2} \frac{\fdag{k}+
\fdag{q_{12}}+m_f} {(k+q_{12})^2-m_f^2 }\gamma^5 \Bigg] \; ,
\end{align}
\begin{align}
T^{\mu_1 \mu_2}_2(q_1,q_2,m_f)= \frac{-\ii}{4m_f} \int \frac{\id^4k}{\ii\pi^2}
\tr \Bigg[ & \frac{\fdag{k}+ m_f}{k^2-m_f^2} \gamma^{\mu_2} \frac{\fdag{k}+
\fdag{q_2}+m_f}{(k+q_2)^2-m_f^2} \gamma^{\mu_1} \frac{\fdag{k}+
\fdag{q_{12}}+m_f} {(k+q_{12})^2-m_f^2 }\gamma^5 \Bigg] \; ,
\end{align}
where $q_1$, $q_2$ are outgoing momenta, $q_{12}=q_1+q_2$ and the overall
factor $-\ii /4m_f$ cancels the explicit mass factor arising from the Dirac
trace. Using the charge conjugation matrix C
\begin{align}
\hat{C} \gamma_{\mu} \hat{C}^{-1}=- \gamma_{\mu}^T \; ,\quad \hat{C} \gamma_5
\hat{C}^{-1}= \gamma_5^T \quad \text{with} \quad
\hat{C}=\gamma^0 \gamma^2, \quad \hat{C}^2= \mathds{1} \; ,
\end{align}
one can derive (Furry's theorem~\cite{Furry}) 
\begin{align}
T^{\mu_1 \mu_2}_1(q_1,q_2,m_f)=T^{\mu_1 \mu_2}_2(q_1,q_2,m_f) \equiv T^{\mu_1
\mu_2}(q_1,q_2,m_f) \; .
\end{align}
Thus, the color structure simplifies to 
\begin{align} 
\tr \big[t^{a_1} t^{a_2}\big] T^{\mu_1 \mu_2}_1 (q_1,q_2,mq) +\tr \big[t^{a_2}
t^{a_1}\big] T^{\mu_1 \mu_2}_2 (q_1,q_2,m_f) & =\delta^{a_1 a_2}T^{\mu_1
\mu_2}(q_1,q_2,m_f) \; .
\end{align}
Evaluation of the Dirac trace yields
\begin{align}
T^{\mu_1 \mu_2}(q_1,q_2,m_f)= \varepsilon^{\mu_1 \mu_2 q_1 q_2}\
C_0(q_1,q_2,m_f) \; .
\end{align}
Here, $C_0$ denotes the scalar three-point function and
$\varepsilon^{\mu \nu q_1 q_2} $ is the totally anti-symmetric tensor
(Levi-Civita symbol), contracted with the gluon momenta $q_1$ and
$q_2$.

\section{Tensor structure of boxes}\label{sec:appB}
\begin{figure}[h!]    
\centering
\includegraphics[scale=0.35,angle=0]{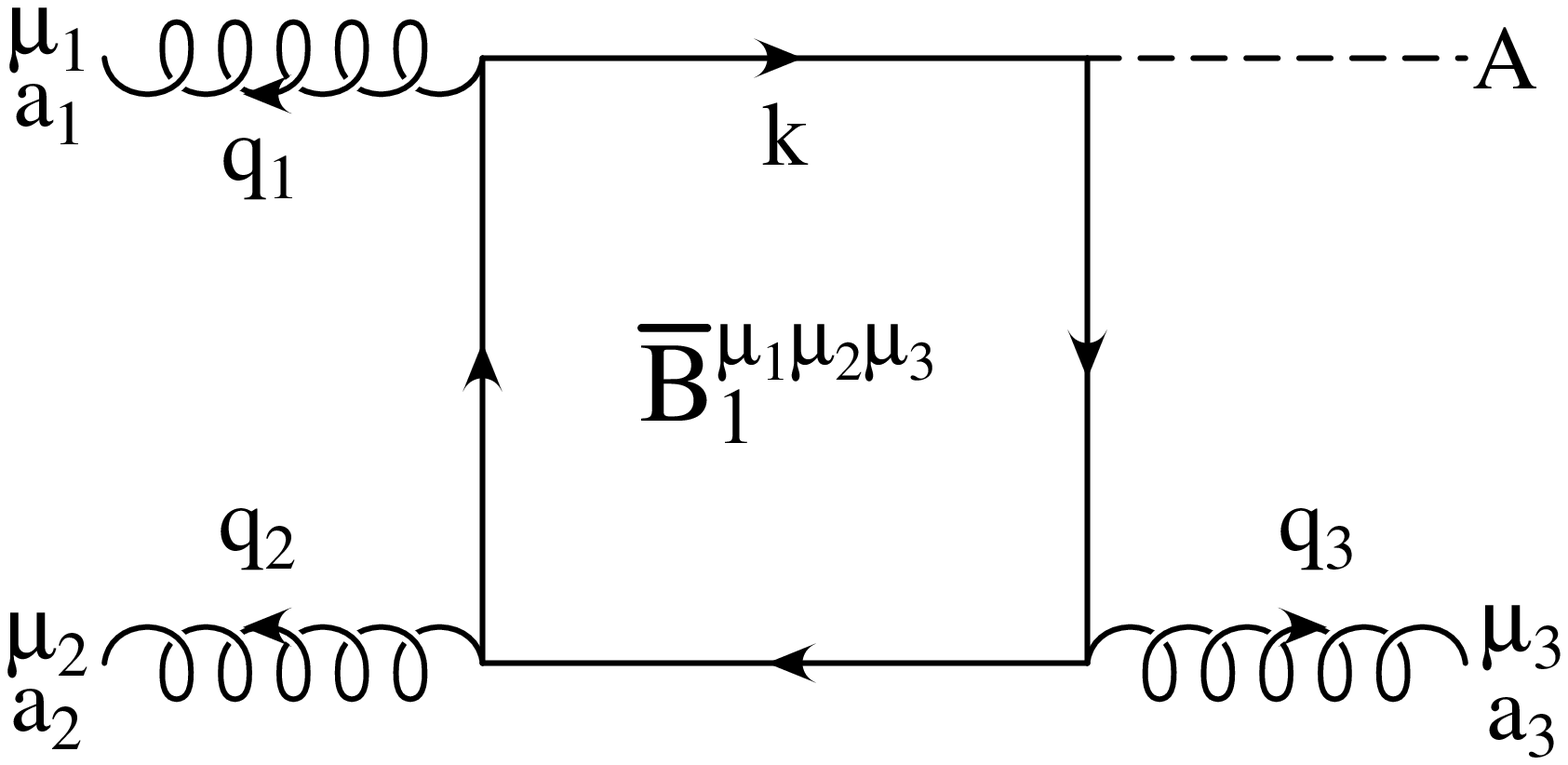} \hspace{1cm}
\includegraphics[scale=0.35,angle=0]{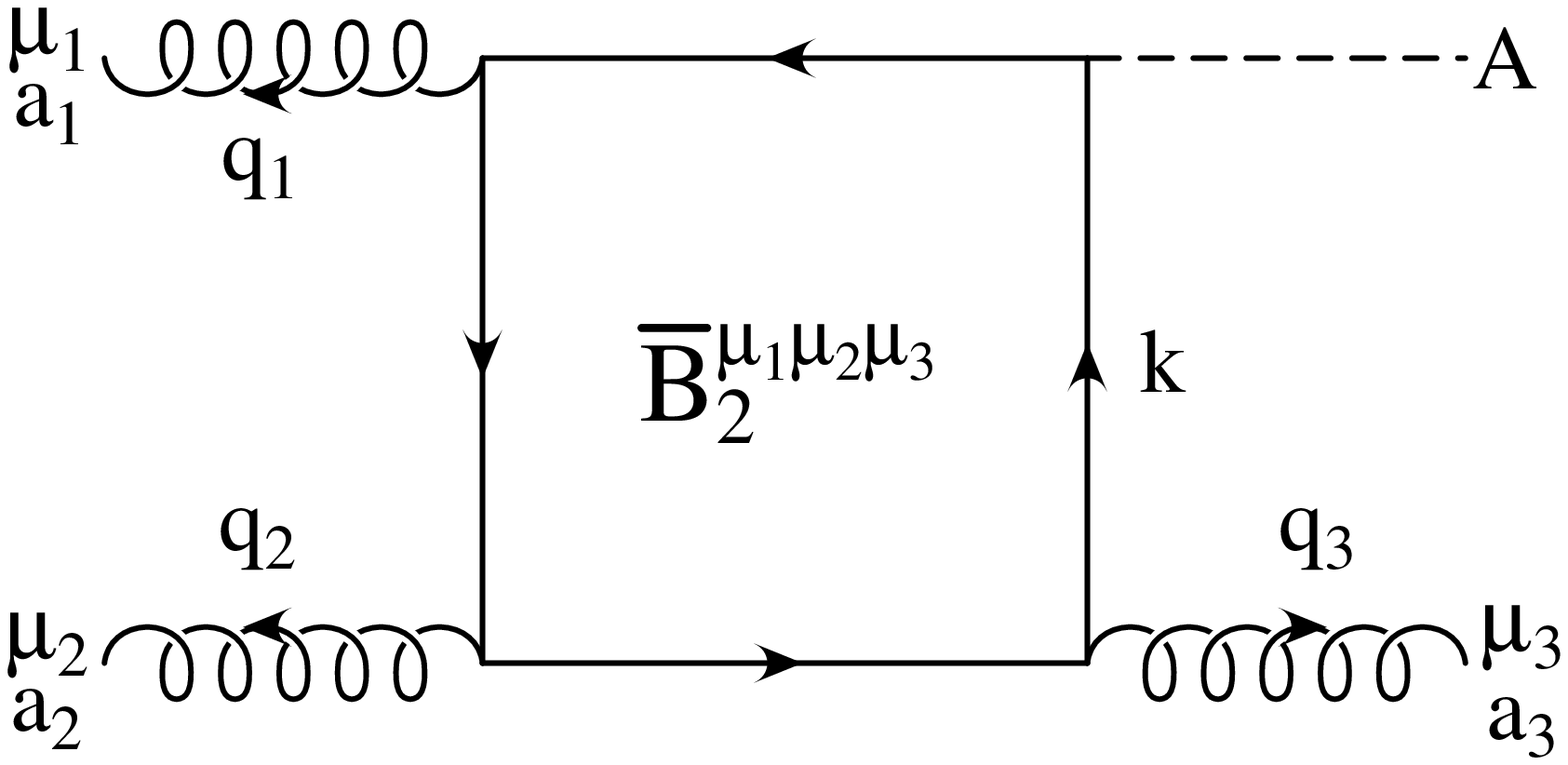}
\ccaption{} 
{ \label{app:box} Two four-point functions connected by charge 
               conjugation.}  
\end{figure} 
The analytic expressions for the charge-conjugated boxes are
\begin{align}
\overline{B}^{\mu_1 \mu_2 \mu_3}_1 (q_1,q_2,q_3,m_f)=
\frac{-\ii}{4m_f} &\int 
\frac{\id^4k}{\ii\pi^2}\tr \Bigg[ \frac{\fdag{k}+
m_f}{k^2-m_f^2} \gamma^{\mu_1}
\frac{\fdag{k}+\fdag{q}_1+m_f}{(k+q_1)^2-m_f^2} \nonumber  \\
\times \quad & \gamma^{\mu_2} \frac{\fdag{k}+ \fdag{q}_{12}+m_f}
{(k+q_{12})^2-m_f^2} \gamma^{\mu_3} \frac{\fdag{k}+ 
\fdag{q}_{123}+m_f} {(k+q_{123})^2-m_f^2} \gamma^5 \Bigg] \; ,
\end{align}
\begin{align}
\overline{B}^{\mu_1 \mu_2 \mu_3}_2 (q_1,q_2,q_3,m_f)=
\frac{-\ii}{4m_f} &\int
\frac{\id^4k}{\ii\pi^2}\tr \Bigg[ \frac{\fdag{k} + m_f}
{k^2-m_f^2} \gamma^{\mu_3} \frac{\fdag{k}+\fdag{q_3}+m_f}{(k+q_3)^2-m_f^2}
\nonumber  \\
\times \quad & \gamma^{\mu_2} \frac{\fdag{k}+ \fdag{q}_{23}+m_f}
{(k+q_{23})^2-m_f^2 } \gamma^{\mu_1} \frac{\fdag{k}
+ \fdag{q}_{123}+m_f} {(k+q_{123})^2-m_f^2 } \gamma^5 \Bigg] \; ,
\end{align}
where $q_1$, $q_2$ and $q_3$ are outgoing momenta, $q_{ij}=q_i+q_j$ and
$q_{ijk}=q_i+q_j+q_k$.
From charge conjugation one gets
\begin{align} \label{furry:box}
\overline{B}^{\mu_1 \mu_2 \mu_3}_1 (q_1,q_2,q_3,m_f)=- \overline{B}^{\mu_1 \mu_2
\mu_3}_2 (q_1,q_2,q_3,m_f) \equiv \overline{B}^{\mu_1 \mu_2 \mu_3}(q_1,q_2,q_3,m_f)
\; .
\end{align}
Further two permutations are obtained by cyclic permutation of
$(1,2,3)$. The color structure for the sum of the two diagrams is
\begin{align}
\tr \big( & t^{a_1}  t^{a_2} t^{a_3} \big) \overline{B}^{\mu_1 \mu_2 \mu_3}_1
(q_1,q_2,q_3,m_f) + \tr \big( t^{a_3} t^{a_2} t^{a_1} \big) \overline{B}^{\mu_1
  \mu_2 \mu_3}_2 (q_1,q_2,q_3,m_f) \nonumber  \\ 
& = \Big[ \tr \big( t^{a_1} t^{a_2} t^{a_3} \big) - \tr \big( t^{a_3} t^{a_2}
t^{a_1} \big) \Big] \overline{B}^{\mu_1 \mu_2 \mu_3} (q_1,q_2,q_3,m_f)
\nonumber \\
&= \frac{\ii}{2}f^{a_1 a_2 a_3} \overline{B}^{\mu_1 \mu_2 \mu_3} (q_1,q_2,q_3,m_f) \; .
\end{align}
The tensor structure of charge-conjugation related box diagrams,
e.g. with gluon permutation $(1,2,3)$, can be written as
\begin{align} \label{boxamp}
&\overline{B}^{\mu_1 \mu_2 \mu_3}(q_1 ,q_2,q_3,m_f) = \Big\{ \
\varepsilon^{\mu_3 q_1 q_2 q_3} \ g^{\mu_1 \mu_2}
-\varepsilon^{\mu_2 q_1 q_2 q_3} \ g^{\mu_1 \mu_3} + 
\varepsilon^{\mu_2 \mu_3 q_2 q_3} \ q_1^{\mu_1} -
\varepsilon^{\mu_2 \mu_3 q_1 q_3} \ q_2^{\mu_1} \nonumber \\
&\quad + \varepsilon^{\mu_2 \mu_3 q_1 q_2} \ q_3^{\mu_1}
+\varepsilon^{\mu_1 q_1 q_2 q_3} \ g^{\mu_2 \mu_3} +
\varepsilon^{\mu_1 \mu_3 q_2 q_3} \ q_1^{\mu_2} - \varepsilon^{\mu_1
  \mu_3 q_1 q_2} \ q_3^{\mu_2}  -
\varepsilon^{\mu_1 \mu_2 q_2 q_3} \ q_1^{\mu_3} \nonumber \\
& \quad + \varepsilon^{\mu_1 \mu_2 q_1q_3} \ q_2^{\mu_3} +
\varepsilon^{\mu_1 \mu_2 \mu_3 q_3} \ g^{\mu_1 \mu_2} -
\varepsilon^{\mu_1 \mu_2 \mu_3 q_2} \ g^{\mu_1 \mu_3} +
\varepsilon^{\mu_1 \mu_2 \mu_3 q_1} \ g^{\mu_2 \mu_3} +
\varepsilon^{\mu_1 \mu_3 q_1 q_3} \big( 2\, q_1^{\mu_2} \nonumber \\
&\quad+ q_2^{\mu_2} \big) + \varepsilon^{\mu_1 \mu_2 q_1 q_2} \ \big[
2\, ( q_1^{\mu_3} + q_2^{\mu_3} ) + q_3^{\mu_3} \big] \Big\} \
D_0(q_1,q_2,q_3,m_f) \nonumber \\
& \quad- \varepsilon^{\mu_1 \mu_2 \mu_3 q_3} \ C_0(q_1+q_2,q_3,m_f) -
\varepsilon^{\mu_1 \mu_2  \mu_3 q_1} \ C_0(q_1,q_2+q_3,m_f) \nonumber \\
& \quad + 2\, \varepsilon^{\mu_2 \mu_3 q_2 q_3} \
D_{\mu_1}(q_1,q_2,q_3,m_f)  + 2\, \varepsilon^{\mu_1 \mu_3 q_1 q_3} \
D_{\mu_2}(q_1,q_2,q_3,m_f) \nonumber \\
&\quad + 2\, \varepsilon^{\mu_1 \mu_2 q_1 q_2} \ D_{\mu_3}(q_1,q_2,q_3,m_f) \; .
\end{align}
The $D_0$ and $D_{\mu}$ are four-point functions. Whereas the former denotes
a scalar function, the latter can be expressed by the usual Passarino-Veltman
decomposition~\cite{Passarino:1978jh} as
\begin{align}
D^{\mu}(q_1,q_2,q_3,m_f) = q_1^\mu \ D_{11} + q_2^\mu \ D_{12} +q_3^\mu \
D_{13} \; .
\end{align}
Note that after contraction with polarization vectors $\epsilon_1^{\mu_1}$,
$\epsilon_2^{\mu_2}$ and quark current $J_{21}^{\mu_3}$, the
expression~\eqref{boxamp} still contains terms with factors $(\epsilon_1 \cdot
q_1),\ (\epsilon_2 \cdot q_2), \ (J_{21} \cdot q_3)$ even though they vanish,
since gluon polarization vectors $\epsilon_i^{\mu} $ and momenta $q_i^{\mu}$
are perpendicular to each other and the quark current $J_{21}$ is
conserved. However, these terms are important for numerical gauge
checks, where the corresponding gluon polarization vector is replaced
by its momentum. Since the virtual gluon has a non-zero
$q_i^2$, these terms give finite contributions.

\section{Tensor structure of pentagons}
\label{sec:appC}
\begin{figure}[ht]     
\centering
\includegraphics[scale=0.35,angle=0]{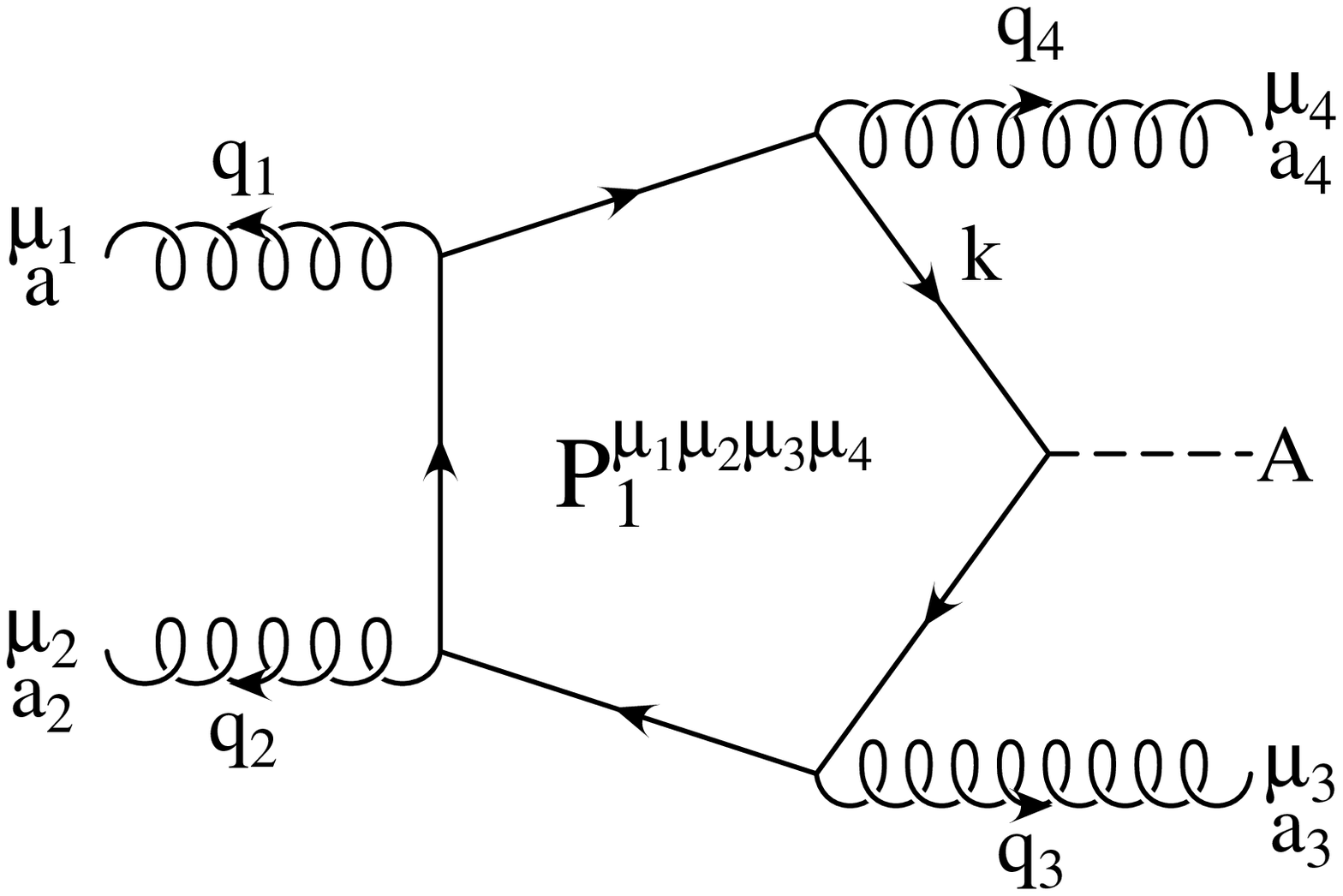} \hspace{1cm}
\includegraphics[scale=0.35,angle=0]{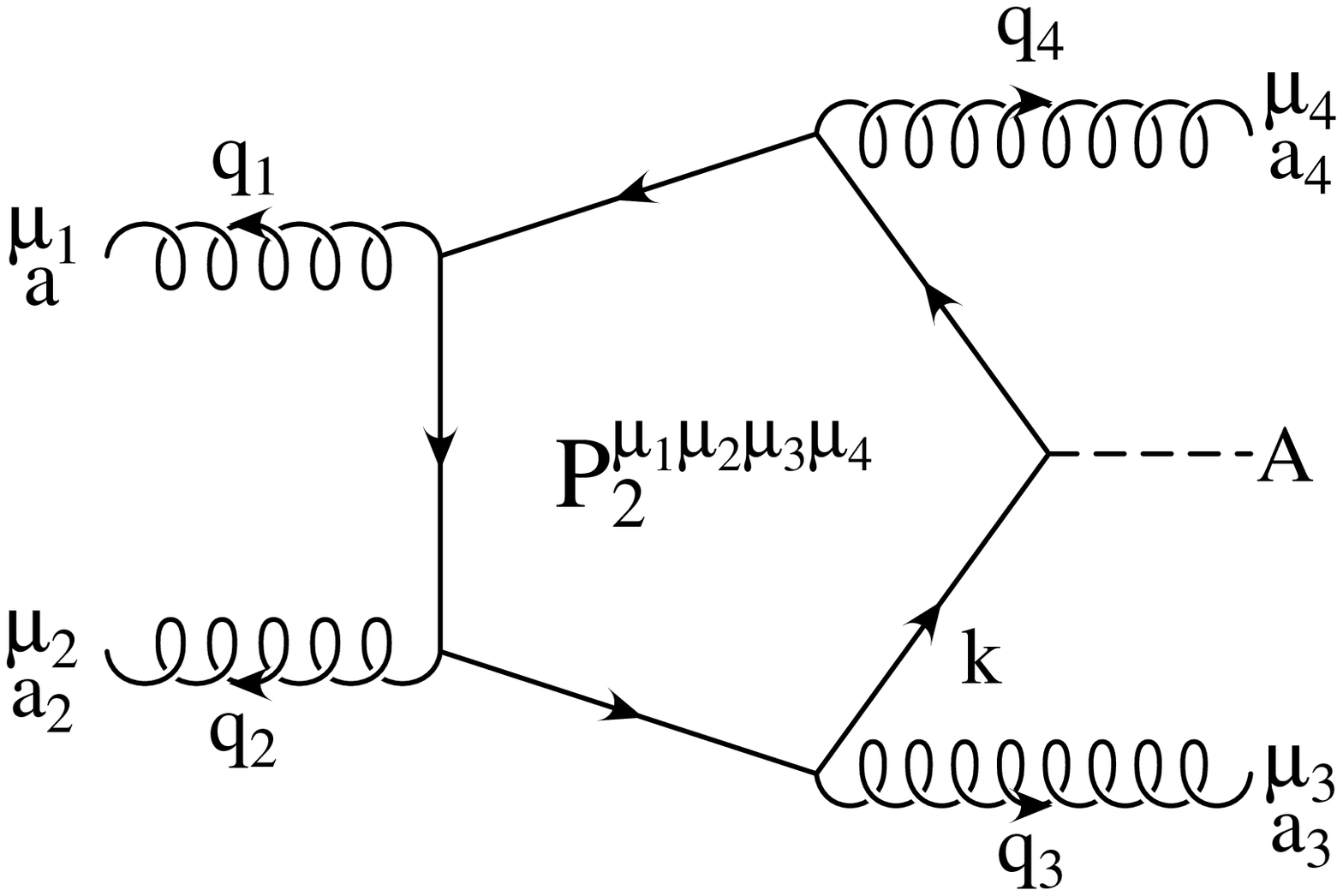}
\ccaption{} 
{ \label{app:pent} Two five-point functions connected by charge 
               conjugation.}  
\end{figure} 
The two five-point functions connected by charge conjugation are defined by the expressions
\begin{align}
 P_1^{\mu_1 \mu_2 \mu_3 \mu_4} ( q_1 ,q_2,& q_3,q_4,m_f) = \frac{-\ii}{4m_f}
\int \frac{\id^4k}{\ii\pi^2}\tr \Bigg[ \frac{\fdag{k}+  m_f}{k^2-m^2_f }
\gamma^{\mu_4} \frac{\fdag{k}+ \fdag{q}_4+m_f}{(k+q_4)^2-m^2_f} \gamma^{\mu_1}
\nonumber \\
& \times \ \frac{\fdag{k}+ \fdag{q}_{14} +m_f} {(k+q_{14})^2-m^2_f }
\gamma^{\mu_2}\frac{\fdag{k}+ \fdag{q}_{124}+m_f} {(k+q_{124})^2-m^2_f}
\gamma^{\mu_3} \frac{\fdag{k}+\fdag{q}_{1234} + m_f} {(k+q_{1234})^2-m^2_f }
\gamma^5 \Bigg] \; ,
\end{align}
\begin{align}
 P_2^{\mu_1 \mu_2 \mu_3 \mu_4} ( q_1 ,q_2,& q_3,q_4,m_f) = \frac{-\ii}{4m_f}
\int \frac{\id^4k}{\ii\pi^2}\tr \Bigg[ \frac{\fdag{k}+  m_f}{k^2-m^2_f }
\gamma^{\mu_3} \frac{\fdag{k}+ \fdag{q}_3+m_f}{(k+q_3)^2-m^2_f} \gamma^{\mu_2}
\nonumber \\
& \times \ \frac{\fdag{k}+ \fdag{q}_{23} +m_f} {(k+q_{23})^2-m^2_f }
\gamma^{\mu_1}\frac{\fdag{k}+ \fdag{q}_{123}+m_f} {(k+q_{123})^2-m^2_f}
\gamma^{\mu_4} \frac{\fdag{k}+\fdag{q}_{1234} + m_f} {(k+q_{1234})^2-m^2_f }
\gamma^5 \Bigg] \; ,
\end{align}
where $q_1$, $q_2$, $q_3$ and $q_4$ are outgoing momenta ($q_{ij}=q_i+q_j$ and
similarly for $q_{ijk}$ and $q_{ijkl}$). The allowed color structures are given 
in~\cite{DelDuca:2001eu}. 
The pentagon
was reduced using the Denner-Dittmaier algorithm~\cite{Denner:2002ii}
and is available as a FORTRAN-subroutine in {\em VBFNLO}~\cite{VBFNLO}. The
full analytic expression in terms of Passarino-Veltman reduction can
be found in~\cite{DA}.

\section*{Acknowledgments}
We would like to thank Gunnar Kl\"amke for helpful discussions, and
Christian Schappacher, for the comparison with the {\em
  FeynArts/FormCalc}~\cite{FA} framework. This research was supported
in part by the Deutsche Forschungsgemeinschaft via the
Sonderforschungsbereich/Transregio SFB/TR- 9 “Computational Particle
Physics”. M.K. gratefully acknowledges support of the
Graduiertenkolleg "High Energy Physics and Particle Astrophysics" and
Landesgraduiertenf\"orderung Baden-W\"urttemberg.
F.C acknowledges partial support by FEDER and Spanish MICINN under grant FPA2008-02878.


\begin{thebibliography}{6}

\bibitem{DelDuca:2001eu}
  V.~Del Duca, W.~Kilgore, C.~Oleari, C.~Schmidt and D.~Zeppenfeld,
  Phys.\ Rev.\ Lett.\  {\bf 87} (2001) 122001
  [arXiv:hep-ph/0105129];
  Nucl.\ Phys.\  B {\bf 616} (2001) 367
  [arXiv:hep-ph/0108030].
\bibitem{CMS}
  G.~L.~Bayatian {\it et al.}  [CMS Collaboration].
  {\em CMS physics: Technical design report}.
\bibitem{ATLAS}
 {\em ATLAS detector and physics performance. Technical design report.
 Vol. 2}, CERN-LHCC-99-15, ATLAS-TDR-15
\bibitem{Djouadi:2005gj}
  A.~Djouadi,
  Phys.\ Rept.\  {\bf 459} (2008) 1
  [arXiv:hep-ph/0503173].
\bibitem{Zeppenfeld:2000td}
  D.~Zeppenfeld, R.~Kinnunen, A.~Nikitenko and E.~Richter-Was,
  Phys.\ Rev.\  D {\bf 62} (2000) 013009
  [arXiv:hep-ph/0002036].
\bibitem{Duhrssen:2004cv}
  M.~Duhrssen et al., 
  Phys.\ Rev.\  D {\bf 70} (2004) 113009
  [arXiv:hep-ph/0406323].
\bibitem{WBFstudies}
  D.~L.~Rainwater and D.~Zeppenfeld,
  JHEP {\bf 9712} (1997) 005
  [arXiv:hep-ph/9712271];
  D.~L.~Rainwater, D.~Zeppenfeld and K.~Hagiwara,
  Phys.\ Rev.\  D {\bf 59} (1999) 014037
  [arXiv:hep-ph/9808468];
  D.~L.~Rainwater and D.~Zeppenfeld,
  Phys.\ Rev.\  D {\bf 60} (1999) 113004
  [Erratum-ibid.\  D {\bf 61} (2000) 099901]
  [arXiv:hep-ph/9906218].
\bibitem{Kauffman:1996ix}
  R.~P.~Kauffman, S.~V.~Desai and D.~Risal,
  Phys.\ Rev.\ D {\bf 55} (1997) 4005
  [Erratum-ibid.\ D {\bf 58} (1998) 119901]
  [arXiv:hep-ph/9610541].
\bibitem{Kauffman:1998yg}
  R.~P.~Kauffman and S.~V.~Desai,
  Phys.\ Rev.\ D {\bf 59} (1999) 057504
  [arXiv:hep-ph/9808286].
\bibitem{Arnold:2008rz}
  K.~Arnold {\it et al.},
  Comput.\ Phys.\ Commun.\  {\bf 180} (2009) 1661
  [arXiv:0811.4559,~hep-ph].
\bibitem{HHG}
  J.~F.~Gunion, H.~E.~Haber, G.~L.~Kane and S.~Dawson, 
{\em The Higgs Hunter's Guide}, Addison-Wesley, Reading (USA), 1990.
\bibitem{Furry}
  W.~H.~Furry,
  Phys.\ Rev.\ {\bf 51} (1936) 125.
\bibitem{Hagiwara:1988pp}
  K.~Hagiwara and D.~Zeppenfeld,
  Nucl.\ Phys.\ B {\bf 313} (1989) 560.
\bibitem{Passarino:1978jh}
  G.~Passarino and M.~J.~G.~Veltman,
  Nucl.\ Phys.\ B {\bf 160} (1979) 151.
\bibitem{FA}
T.~Hahn, Comput. Phys. Commun. 140 (2001) 418 [arXiv:hep-ph/0012260];
T.~Hahn and C. Schappacher, Comput. Phys. Commun. 143 (2002) 54
[hep-ph/0105349];
T.~Hahn, M. Perez-Victoria, Comput. Phys. Commun. 118 (1999) 153
[hep-ph/9807565]
\bibitem{DA}
M.~Kubocz, {\em Diploma thesis} (german), Institut f\"ur Theoretische Physik,
Universit\"at Karlsruhe, 2006, \newline
http://www-itp.particle.uni-karlsruhe.de/diplomatheses.de.shtml
\bibitem{VBFNLO}
The VBFNLO code can be obtained from 
http://www-itp.particle.uni-karlsruhe.de/vbfnlo/
\bibitem{Denner:2002ii}
  A.~Denner and S.~Dittmaier,
  Nucl.\ Phys.\  B {\bf 658} (2003) 175
  [arXiv:hep-ph/0212259];
  Nucl.\ Phys.\  B {\bf 734} (2006) 62
  [arXiv:hep-ph/0509141].
\bibitem{CTEQ6}
 J.~Pumplin, D.~R.~Stump, J.~Huston, H.~L.~Lai, P.~Nadolsky and W.~K.~Tung,
  JHEP {\bf 0207} (2002) 012
  [arXiv:hep-ph/0201195].
\bibitem{Spira:1997dg}
  M.~Spira,
  Fortsch.\ Phys.\  {\bf 46} (1998) 203
  [arXiv:hep-ph/9705337].
\bibitem{Vermaseren:1997fq}
  J.~A.~M.~Vermaseren, S.~A.~Larin and T.~van Ritbergen,
  Phys.\ Lett.\  B {\bf 405} (1997) 327
  [arXiv:hep-ph/9703284].
\bibitem{Plehn:2001nj}
  T.~Plehn, D.~L.~Rainwater and D.~Zeppenfeld,
  Phys.\ Rev.\ Lett.\  {\bf 88} (2002) 051801
  [arXiv:hep-ph/0105325].
\bibitem{Hankele:2006ja}
  V.~Hankele, G.~Klamke and D.~Zeppenfeld,
  ``Higgs + 2 jets as a probe for CP properties,''
  [arXiv:hep-ph/0605117].
\bibitem{Hankele:2006ma}
  V.~Hankele, G.~Klamke, D.~Zeppenfeld and T.~Figy,
  Phys.\ Rev.\  D {\bf 74} (2006) 095001
  [arXiv:hep-ph/0609075].
\bibitem{Klamke:2007pn}
  G.~Klamke and D.~Zeppenfeld,
  ``Hjj production: Signals and CP measurements,''
  [arXiv:0705.2983,~hep-ph].
\bibitem{Klamke:2007cu}
  G.~Klamke and D.~Zeppenfeld,
  JHEP {\bf 0704} (2007) 052
  [arXiv:hep-ph/0703202].
\bibitem{Campbell:2006xx}
  J.~M.~Campbell, R.~K.~Ellis and G.~Zanderighi,
  JHEP {\bf 0610} (2006) 028
  [arXiv:hep-ph/0608194].
\bibitem{Boos:2001cv}
  E.~Boos {\it et al.},
  [arXiv:hep-ph/0109068];
  J.~Alwall {\it et al.},
  Comput.\ Phys.\ Commun.\  {\bf 176} (2007) 300
  [arXiv:hep-ph/0609017].

\end{thebibliography}
\end{document}